\pdfoutput=1
\documentclass[a4paper,11pt]{article}
\usepackage{jheppub} 
\usepackage{comment}
\usepackage{amsmath,amssymb,amsfonts,mathtools}
\usepackage[toc,page]{appendix}
\usepackage{hyperref}
\usepackage{physics}
\usepackage{float}
\usepackage{slashed}
\usepackage{multirow}
\usepackage[dvipsnames]{xcolor}
\usepackage{tikz}
\usepackage{subfigure}
\usetikzlibrary{decorations.markings, arrows.meta, decorations.pathmorphing}

\renewcommand{\dd}{\mathrm{d}}
\newcommand{\mt}[1]{\textrm{\scriptsize #1}}
\renewcommand{\l}{\lambda}

\def\Nc{N_\mt{c}}
\def\Nf{N_\mt{f}}
\def\Vf{V_\mt{f}}
\def\Vg{V_\mt{g}}
\def\rh{r_\mt{h}}
\def\Mp{M_\mt{p}}

\def\nB{n_\mt{B}}
\def\nS{n_\mt{S}}

\renewcommand{\dd}{\mathrm{d}}

\newcommand{\ie}{{\emph{i.e.}}}
\newcommand{\eg}{{\emph{e.g.}}}

\def\muB{\mu_\mt{B}}
\def\muQ{\mu_\mt{Q}}
\def\muS{\mu_\mt{S}}

\def\nB{n_\mt{B}}
\def\nS{n_\mt{S}}
\def\nQ{n_\mt{Q}}

\def\chitwo{\chi_2^\mt{B}}

\def\chifour{\chi_4^\mt{B}}

\def\chitwoT{\tilde{\chi}_2^\mt{B}}

\def\chifourT{\tilde{\chi}_4^\mt{B}}

\def\chinT{\tilde{\chi}_n^\mt{B}}

\newcommand{\be}{\begin{equation}}
\newcommand{\ee}{\end{equation}}
\newcommand{\bea}{\begin{eqnarray}}
\newcommand{\eea}{\end{eqnarray}}

\preprint{\text{HIP-2026-19/TH}}

\title{The holographic QCD critical point is sensitive to quark flavors
}

\author[a]{Niko Jokela,} 
\author[b]{Matti J\"arvinen,}
\author[c]{Toshali Mitra,}
\author[a]{and Aleksi Piispa} 
\affiliation[a]{Department of Physics and Helsinki Institute of Physics, \\
P.O. Box 64, FI-00014, University of Helsinki, Finland}
\affiliation[b]{Institute of Theoretical Physics, Chinese Academy of Sciences, Beijing 100190, China}
\affiliation[c]{Institute for Theoretical Physics, University of Heidelberg, D-69120 Heidelberg, Germany}

\emailAdd{t.mitra@thphys.uni-heidelberg.de}
\emailAdd{niko.jokela@helsinki.fi}
\emailAdd{mattijarvinen@itp.ac.cn}
\emailAdd{aleksi.piispa@helsinki.fi}

\abstract{We study the phase structure of dense QCD matter by applying the flavor-dependent holographic V-QCD model to two distinct physical environments: charge-neu\-tral, $\beta$-equi\-lib\-ra\-ted matter, and strangeness-neutral matter with a fixed charge-to-baryon number ratio, $\nQ/\nB = 0.4$, relevant for heavy-ion collisions. The quark sector of the model incorporates the realistic mass hierarchy of two massless light quarks and a massive strange quark. We find that the resulting phase diagram depends sensitively on both the procedure used to tune the model against lattice QCD thermodynamics and the choice of physical environment. Using our preferred fitting procedure, we find that 
$\beta$-equilibrated matter exhibits a first-order phase transition terminating at a critical endpoint located at significantly lower densities than in earlier holographic studies. However, this phase transition disappears entirely under heavy-ion conditions. This pronounced environmental dependence may help explain why recent net-proton cumulant measurements from Phase II of the RHIC Beam Energy Scan have not yet revealed clear non-monotonic fluctuation signatures. We calculate higher-order baryon number cumulant ratios along the chemical freeze-out curve to quantitatively compare the model predictions with experimental data. Finally, we also present global phase diagrams obtained by matching V-QCD to Hadron Resonance Gas models implemented in the \texttt{Thermal-FIST} package, and argue that this comparison further supports the conclusions drawn from the V-QCD model alone.}

\begin{document}

\maketitle
\flushbottom
\setcounter{page}{2}

\clearpage

\section{Introduction}\label{sec:intro}

Mapping the phase diagram of Quantum Chromodynamics (QCD) at nonzero temperature and baryon chemical potential is a central objective in the study of strongly interacting matter.  While high-energy experiments such as the second phase of the Beam Energy Scan (BES-II) at RHIC probe the hot and dense matter created in relativistic heavy-ion collisions (HIC), astrophysical observations of neutron stars constrain the equation of state (EoS) at low temperatures and extreme densities. Connecting these two regimes theoretically requires non-perturbative methods. However, the standard first-principles non-perturbative method, lattice QCD, is restricted by the sign problem at nonzero baryon chemical potential~\cite{deForcrand:2009zkb}. As a result, investigating the phase structure of strongly coupled quark matter at high density relies heavily on modeling. Over the past two decades, the gauge/gravity duality~\cite{Ramallo:2013bua}, or simply holography, has become a standard alternative non-perturbative approach, mapping the complex dynamics of four-dimensional gauge theories onto higher-dimensional classical gravity backgrounds.

In the context of bottom-up holography,\footnote{While this work focuses exclusively on bottom-up phenomenology, it is important to acknowledge the top-down approach to holographic QCD
~\cite{Kovensky:2026zdd}. 
The top-down framework is highly appealing because its microscopic degrees of freedom and string-theoretic origins are in principle fully under theoretical control. However, because its structure is strictly governed by the underlying fundamental theory, it cannot be systematically tuned to fit lattice or experimental thermodynamic data. Therefore, our present data-driven analysis necessarily relies on the more flexible bottom-up approach.} the standard methodology relies on matching a five-dimensional holographic gravity action to lattice QCD data at zero or small density, and subsequently using the resulting gravitational background to extrapolate the EoS into the high-density regime. This extrapolation relies heavily on the structural choices built into the holographic action, leading to different phenomenological branches that handle the vacuum and thermodynamic properties in distinct ways. 

The research follows two main branches. The first branch consists of models based on the improved holographic QCD framework (IHQCD)~\cite{Gursoy:2007cb,Gursoy:2007er}. Such models have a confining vacuum and (unlike real QCD) contain a first-order confinement-deconfinement transition even at zero density. However, given that they are confining,  apart from the thermodynamics they can also be used to model the spectrum of QCD.  
The second branch, following original works~\cite{Gubser:2008ny,DeWolfe:2010he,DeWolfe:2011ts}, uses models that are not confining, and therefore cannot describe the hadron spectrum of QCD. However, since such models do not show a first-order phase transition (FOPT) at zero density, they offer a simpler and more direct way of fitting the lattice QCD data. This branch of work typically uses Einstein--Maxwell-dilaton (EMD) models and is often termed  ``black hole engineering.'' Both the IHQCD and EMD branches can be extended to include additional sectors, \eg, description of chiral symmetry breaking~\cite{Jarvinen:2011qe,Shen:2025zkj}. 

The main motivation for the works on the EMD models has been the study of the possible critical point of the QCD phase diagram. Already the early studies~\cite{DeWolfe:2010he,DeWolfe:2011ts} demonstrated the presence of a critical point at finite density in such holographic models, and these studies have been refined later on~\cite{Jokela:2024xgz,Li:2025lmp,Shen:2025zkj}. While a critical point therefore appears naturally in the EMD models, in the extension of the IHQCD framework to finite density after implementing quarks (\ie, using the V-QCD model~\cite{Jarvinen:2011qe}), studying the critical point does not seem to be possible at first sight: The confined and deconfined phases cannot be smoothly connected, so the deconfinement phase transition cannot end in a critical point. However, it was recently pointed out that this is not necessarily true, if one goes beyond pure holographic gravity description of the phase diagram. That is, by comparing the results of V-QCD in the deconfined quark-gluon plasma phase to a simple hadron resonance gas model of the confined phase, a critical point could be identified~\cite{Demircik:2021zll,Ecker:2025vnb}. Interestingly, the location of the critical point found in this construction closely agrees (within $\sim$10\% precision in temperature and baryon chemical potential) with the location found in the most recent EMD studies.

Within the EMD program, there have also been attempts to include the flavor degrees of freedom. However, most of the literature only includes the effect of the flavors through fitting lattice data ($\Nf=2+1$ or $\Nf=2+1+1$) where flavor quark effects are included, but restricting to a model with a single chemical potential~\cite{Knaute:2017opk,Critelli:2017oub,Grefa:2021qvt,Cai:2022omk,He:2023ado,Hippert:2023bel,Zhao:2023gur,Fu:2024nmw,Chen:2024mmd,Cai:2024eqa,Zhu:2025gxo,Tong:2026dwb}. Additional charge and strangeness chemical potentials and gauge fields were introduced in~\cite{Rougemont:2017tlu} but as probes, so that the critical point is unaffected. A related extension of the EMD which introduces a separate gauge field for each quark flavor with different couplings (but no chiral symmetry breaking) was studied recently~\cite{Li:2026lxx}. While there is work on adding flavor-independent scalar fields in the EMD setup~\cite{Liu:2023pbt},
flavor-dependent scalars on top of the EMD framework have also been introduced as probes~\cite{Cai:2022omk} and, recently, taking into account backreaction~\cite{Shen:2025zkj,Deng:2026aht}. However, to our knowledge, no generalization to fully flavored Einstein--Yang--Mills-dilaton model with chiral symmetry breaking, has been established so far. 

Both the EMD and IHQCD (extended to V-QCD) modeling carried out in the literature suffer from potential weaknesses or issues that require more attention. Such issues include the following:
\begin{itemize}
    \item[I] The procedure of treating the flavors and fitting of lattice QCD data. Namely, while there are a few attempts to introduce quark flavors in some way in the holographic models, most of the studies simply ignore the flavor structure. The most important shortcoming seems to be that most existing studies constrain the density dependence of the holographic model only by tuning the holographic model to fit the zero-density baryon number susceptibility $\chitwo$ (and potentially higher-order susceptibilities of the baryon number~\cite{Jokela:2024xgz}). Treating dense matter through the lens of a single, global baryon charge obscures the underlying flavor structure of real-world QCD: It is known from the lattice studies that the susceptibilities of QCD at zero density do have nontrivial flavor structure~\cite{Borsanyi:2011sw,HotQCD:2012fhj}. This structure can be characterized in terms of the quark susceptibility matrix $\chi_{ij}$, with $i,j = u,d,s$. Lattice data indicates that the susceptibilities for light quarks and the strange quark differ significantly, which is missed by holographic models unless the flavor structure is implemented properly. Moreover, as it turns out, practically none of the holographic models considered in the literature can even in principle fit all the lattice data faithfully even if flavors are introduced. This is the case because the models predict zero off-diagonal quark susceptibilities ($\chi_{ij}=0$ with $i\ne j$) at zero density. Lattice data shows that the off-diagonal terms are indeed negligible at high temperatures, but the light-quark-strange-quark cross susceptibility becomes significant near the crossover at $T \approx 150$~MeV. Note that when the baryon number susceptibility $\chitwo$ is expressed in terms of the quark susceptibilities, it does depend on this off-diagonal term. The majority of fits carried out in the literature are in models which predict zero off-diagonal susceptibilities but use $\chitwo$, and are therefore not strictly consistent with holography. 
\item [II] Instability of the models at nonzero density~\cite{Demircik:2024aig,CruzRojas:2024igr}. That is, when holographic QCD models are fitted to lattice thermodynamics using only the total baryon charge, an inhomogeneous instability appears already at moderate densities. This instability belongs to the class known as the Nakamura--Ooguri--Park instabilities~\cite{Nakamura:2009tf}. It is driven by Chern--Simons terms~\cite{Domokos:2007kt,Nakamura:2009tf,Bergman:2011rf}, whose presence in holographic QCD is required in order to match with the global flavor anomalies of QCD. In~\cite{Demircik:2024aig} it was shown that the instability is universal: the extent of the unstable region depends only weakly on the choice of the action or details of the fit to lattice data, assuming a simple fit to $\chitwo$ only. Importantly, the instability cloaks the critical point, \ie, the instability sets in only at nonzero density but well before reaching the critical point. Therefore, the critical point constructed by using only homogeneous solutions will not be present in the final phase diagram of the model. The model dependence reported in~\cite{Demircik:2024aig} is so weak that as far as we can see this observation applies to all holographic analyses of the QCD critical point in the literature, with the understanding that the models are extended to include the full chiral symmetry of QCD and Chern-Simons terms implementing QCD anomalies, as required by the holographic dictionary. 
However, if one turns on flavors in the holographic model, the onset and extent of this instability region may depend strongly on the mass of the strange quark~\cite{Demircik:2024aig}. This motivates a systematic flavor-dependent analysis, highlighting that generic black hole engineering is no longer sufficient. To properly capture the phase structure and stability of dense quark matter, a holographic model must track realistic, flavor-dependent physics. Consequently, whether the specific critical endpoint found in our flavored setup is cloaked by such an instability remains an open problem.
\item[III]The holographic implementation of the confined phase. It is well known that the pressure of real QCD in the confined region is given by a weakly interacting gas of hadrons, with pions dominating at lowest temperatures but other mesons and even baryons becoming important near the crossover. The meson pressure is suppressed with respect to the pressure of the deconfined quark-gluon plasma by $1/\Nc^2$, but this suppression is compensated by the relatively large number 
mesons states in real QCD, so that the hadron gas pressure cannot be neglected. Moreover, as also seen from the large $\Nc$ counting~\cite{tHooft:1973alw}, on the gravity side the hadron gas is not captured by classical gravity but should arise from string loops. This casts serious doubts on the validity of a purely gravitational approach to model the transition and the critical point. There have been attempts~\cite{Alho:2015zua,Demircik:2021zll,Yang:2026brr} in the literature where one uses holography only on the deconfined side and compares to a phenomenological or hadron resonance gas model on the confined side, but this necessarily means reduced control over the properties of the transition.
\end{itemize}

In this article, we focus on the first of these three points, \ie, the implementation of flavors, comparison to lattice data, and its implications for the phase diagram. To this end, we carry out a detailed fit to lattice data in a holographic model fully implementing flavor dependence, and study implications to the phase diagram and the critical point.
We will also partially address the point III, \ie, the problem of properly implementing the confined phase of QCD in holography, by comparing our results to hadron resonance gas models. Studying inhomogeneous instability (point II) is postponed to future work. Note that addressing point I means that we will be moving beyond a simple fit to $\chitwo$ only, which in turn means that the extent of the instability (and whether it affects the critical point) is an open question.

We use the flavor-dependent variant of the holographic V-QCD model~\cite{CruzRojas:2024etx,Jarvinen:2025mgj}, which consistently couples the improved holographic QCD gluon sector with a tachyonic D-brane picture. Rather than forcing a fit to baryon susceptibilities, the implementation~\cite{Jarvinen:2025mgj} handles the quark sector by explicitly incorporating a physical mass hierarchy: the up and down quarks are kept massless, while the strange quark is massive. However, even the flavor-dependent variant of the model still predicts zero off-diagonal susceptibilities at zero density, so that completely consistent fit to the data is not possible. 
Here, following~\cite{Jarvinen:2025mgj}, we choose only to fit the diagonal components of the quark susceptibility matrix to lattice results. That is, the model parameters are matched directly to lattice data for the individual light and strange quark susceptibilities from zero-density lattice QCD~\cite{Borsanyi:2011sw,HotQCD:2012fhj}. Note that this means that our results for the baryon number susceptibility $\chitwo$ are in slight disagreement with lattice data at zero density near the crossover, because (as we already pointed out above) the baryon number susceptibility depends on the off-diagonal components of the susceptibility matrix.

While the flavor-dependent V-QCD setup of~\cite{Jarvinen:2025mgj} provides the necessary physical starting point for studying flavor effects in V-QCD, its implications for the global phase diagram were not yet mapped out in~\cite{Jarvinen:2025mgj}: only results at zero density or zero temperature were published. In this article, we start the systematic exploration of the phase diagram as a function of temperature and baryon number chemical potential. We focus on 
the regime relevant for HIC experiments. 

Performing a systematic scan of the solutions over the relevant parameter space, we uncover a FOPT between deconfined phases ending in a critical point on the  phase diagram drawn as a function of the temperature and the baryon number chemical potential. In earlier analyses of V-QCD where the strange quark mass was set to zero~\cite{Jokela:2018ers} and which relied on fits to global baryon susceptibility data, such a transition existed, but was thermodynamically subdominant. Now, matching the model directly to $\Nf=2+1$ flavor data makes this FOPT physical and brings it down to lower chemical potentials.  This means that  our result is also drastically different from earlier EMD studies. 
A key result of this paper is that the physical relevance of this transition is sensitive to the specific conditions imposed by the environment. To be precise, the FOPT and its terminating critical endpoint (CEP) are visible when the V-QCD model is evaluated under the charge-neutral, $\beta$-equilibrated conditions (neglecting leptons), similar to those considered in earlier works where the effects of flavors were not fully addressed. These conditions mean that the system is long-lived enough for weak interactions to establish equilibrium between the quark flavors, enforcing the equality of the down and strange quark chemical potentials ($\mu_d=\mu_s$). This, coupled with the requirement of a vanishing net electric charge density ($\nQ=0$), then dictates the thermodynamic state of the matter. Conversely, when we apply the exact same model to the environment of relativistic HIC, imposing strict strangeness neutrality ($n_\mathrm{S}=0$) and a fixed charge-to-baryon ratio ($\nQ/\nB=0.4$), the FOPT between deconfined phases, and its associated CEP, vanish entirely from the phase diagram.

This striking environmental dependence offers a potential explanation for the recent experimental results from BES-II. High-statistics measurements of net-proton fluctuations by the STAR collaboration~\cite{STAR:2025ydk} have revealed no sign of a two-component structure~\cite{Bzdak:2018uhv,Bzdak:2018axe} or the characteristic non-monotonic signatures~\cite{Stephanov:2011pb} that would indicate a FOPT and a CEP in the heavy-ion context. While it is still possible that this is due to limited accuracy of the results, our holographic analysis suggests an alternative explanation for this absence: the physical constraints of a heavy-ion collision actively alter the structure of the phase diagram. We recognize that various functional methods~\cite{Fu:2019hdw,Fischer:2018sdj}, effective models~\cite{Fukushima:2008wg}, and recent lattice extrapolations~\cite{Basar:2023nkp,Shah:2024img,Shah:2026mzu} predict a finite-density critical point. While state-of-the-art functional renormalization group (fRG) and Dyson-Schwinger studies of QCD have recently begun to map the phase diagram under enforced heavy-ion conditions~\cite{Wen:2019ruz,Gunkel:2021oya,Fu:2026qnl}, many non-holographic calculations remain restricted by truncated zero-density expansions or simplified flavor conditions.
Our observation that environmental constraints themselves may wash out the phase transition provides a clear motivation for future theoretical studies: Critically reevaluating effective models  and extrapolations under heavy-ion conditions will be crucial to determine if the vanishing of the CEP is a universal feature of the QCD phase diagram.

We also investigate the higher-order baryon number fluctuations, usually discussed in terms of the cumulants $C_n$ (where $n \ge 1$) of the net-proton number fluctuations. We compute them in the HIC environment through their relation to the equation of state.  Despite the surprising results for the location of the critical point, we find good agreement with heavy-ion data for the ratios $C_2/C_1$ and $C_4/C_2$. However, there is a significant mismatch between the model and data for $C_3/C_2$. We argue that this mismatch originates from the issue of failing to fit the off-diagonal quark susceptibilities.

As we discussed above, classical gravity analysis of purely holographic backgrounds cannot give a faithful description of the confined hadronic phase in QCD. 
However, in order to introduce hadron gas contributions and construct a physically complete picture of dense QCD phase diagram, the high-temperature deconfined EoS obtained from V-QCD can be matched with a realistic low-temperature hadronic phase~\cite{Alho:2015zua,Jokela:2020piw}. 
We achieve the matching by comparing our holographic results to different variants of the Hadron Resonance Gas (HRG) model, specifically the Excluded Volume (EVX)~\cite{Rischke:1991ke,Vovchenko:2020lju} and the quantum van der Waals (vdW) formulations~\cite{Vovchenko:2015vxa,Vovchenko:2016rkn}. The corresponding hadronic thermodynamic quantities are generated using the \texttt{Thermal-FIST} package~\cite{Vovchenko:2019pjl}, allowing us to locate the matching boundaries by identifying where the pressures of the two independent frameworks intersect. We shall argue that this comparison works well, \ie, a relatively good agreement between the models is found for a wide range of chemical potentials, which adds to the credibility of both the HRG and holographic approaches near the matching boundaries.

The matching procedure between the HRG and V-QCD models highlights a subtle but critical thermodynamic problem inherent to heavy-ion physics context. When analyzing a first-order phase transition in a multi-charge system under strict density conditions (\eg, $\nS=0$ and $\nQ/\nB=0.4$), there is no clean, mathematically consistent thermodynamical ensemble defined in terms of the baryon chemical potential $\muB$ that satisfies these local density conditions across a putative phase boundary. Forcing the conditions to hold identically in both independent phases leads to a splitting of the phase transition line in the $(\muB,T)$-plane because the individual charge chemical potentials fail to match. Moreover, a completely precise comparison of phase dominance would require a non-congruent treatment allowing charge redistribution, leading to a mixed phase where local charge densities are allowed to change~\cite{congruentdefinition}.\footnote{Note that this is true even for the $\beta$-equilibrium setup, since the requirement of vanishing charge density means that the charge chemical potential is not matched at transitions.} In this paper, we focus on the congruent setup.

It is important to emphasize that the results presented in this paper do not represent the final picture of dense holographic QCD or the V-QCD model. While the environmental dependence of the phase diagram offers valuable physical insights, plenty of work remains. In this work, we ignored both the  holographic nuclear matter phase~\cite{Ishii:2019gta} in V-QCD and the hairy black hole branches dual to chirally broken deconfined phases~\cite{Alho:2013hsa,Jarvinen:2025mgj}. The former would be relevant for studies of matter only in the higher density edge of the regime studied in this article. The latter is expected to be relevant at temperatures well below the crossover temperature in QCD, so we expect it to be irrelevant for the high-temperature physics, which is our focus here.
Moreover, refining the holographic framework to systematically capture missing physical elements, such as complete cross-flavor correlations, mixed-phase thermodynamics, or spatially inhomogeneous phases, is an essential next step toward a more realistic description of dense quark matter.

The remainder of this paper is organized as follows. In Section~\ref{sec:setup}, we introduce the holographic V-QCD model and review its salient properties necessary for the thermodynamic analysis in this paper. In Section~\ref{sec:thermo}, we review the thermodynamical setup of the model. In Section~\ref{sec:betaequil}, we explore the phase structure and CEP of the V-QCD model under both $\beta$-equilibrated and heavy-ion conditions, and we compute higher-order baryon number cumulants to compare with experimental data from the RHIC BES program. In Section~\ref{sec:heavyion}, we extend this framework by matching the high-temperature holographic quark-gluon plasma to HRG models, constructing a phase diagram that incorporates the confined hadronic phase. Finally, in Section~\ref{sec:discussion}, we summarize our findings, discuss the limitations of the current framework, and outline directions for future work. Supplementary details regarding the near-boundary asymptotics, numerical methods, and parameter fitting are provided in the appendices.

\section{Holographic model}\label{sec:setup}

In this section we review the generalized flavor-dependent V-QCD model, following \cite{Jarvinen:2025mgj}. V-QCD is a bottom-up holographic model inspired by the five-dimensional non-critical string theory~\cite{Gursoy:2007cb,Jarvinen:2011qe} and formulated in the Veneziano limit of QCD~\cite{Veneziano:1976wm},  where the number of colors $\Nc$ and the number of flavors $\Nf$ are taken to infinity keeping their ratio $x\equiv \Nf/\Nc$ fixed. 

As we shall show below, the V-QCD approach introduces various parameters and potentials into the action, which are determined phenomenologically by comparison with the lattice QCD results. In particular, instead of strict Veneziano limit, one in practice considers $\Nc=3$ and $\Nf=2$ or $\Nf=3$ when comparing to QCD. In this article, we use the flavor-dependent V-QCD model by typically setting $\Nf=2+1$, \ie, two light quarks and a massive strange quark. One could in principle keep working in the Veneziano limit by dividing the large number of flavors into three groups corresponding to the physical quark flavors, while keeping also $\Nc$ to be large. However, we choose to simply set $\Nf=3=\Nc$ from the start and neglect corrections in $1/\Nc$ or $1/\Nf$.\footnote{The $1/\Nc$ effects in thermodynamics have been seen to be small by lattice analysis in pure Yang-Mills theory \cite{Panero:2009tv,Rindlisbacher:2025dqw}. While similar analysis in full QCD is challenging, the Yang-Mills result suggests that including the $1/\Nc$ and $1/\Nf$ corrections for full QCD only through the use of data at $\Nc=\Nf=3$ is a reasonable approximation.}

The V-QCD model~\cite{Jarvinen:2011qe} consists of a gluon sector, described on the gravity side by the IHQCD model~\cite{Gursoy:2007cb,Gursoy:2007er,Gursoy:2010fj}, and a quark sector arising from a tachyonic D-brane picture~\cite{Bigazzi:2005md,Casero:2007ae,Bergman:2007pm,Dhar:2007bz,Jokela:2009tk}. 
The relevant five-dimensional fields are dual to the relevant and marginal operators in QCD. The corresponding V-QCD dictionary is summarized in Table~\ref{tab:vqcd_dictionary}. Here, $g_{MN}$ denotes the five-dimensional bulk metric, whose boundary value gives the field-theory metric $\eta_{\mu\nu}$ and which is dual to the energy-momentum tensor $T_{\mu\nu}^\mathrm{em}$.\footnote{Our conventions are such that the capital Latin indices $M,N,\ldots$ run over all five dimensions whereas Greek indices $\mu,\nu,\ldots$ only run over the spacetime dimensions. As usual, the dictionary is simple if one assumes the Fefferman--Graham and radial gauges, $g_{r\mu}=0= A_r^{ij}$  with $r$ denoting the holographic coordinate.} The scalar field $\lambda = e^{\phi}$ is identified with the running ’t Hooft coupling near the boundary, $\lambda_t = g_\mt{YM}^2 \Nc$, and is dual to the gluonic operator $\mathrm{tr}G_{\mu\nu}G^{\mu\nu}$, where $G_{\mu\nu}$ is the gluon field-strength tensor. The field $T_{ij}$ is the tachyon, represented as a complex matrix in flavor space with $i,j = 1,\ldots ,\Nf$, and is dual to the quark bilinear $\bar \psi^j \psi^i$. Its source is the complex quark-mass matrix $M_{ij}$. Finally, $A^{ij}_{M}$ denotes the vector gauge field in the flavor sector. Its boundary value acts as an external source $A_\mu^{({\rm ext})ij}$ for the flavor current $\bar{\psi}^i \gamma_\mu \psi^j$.

\begin{table}[htb]
\centering
\begin{tabular}{ |p{3cm}||p{3cm}||p{3cm}| }
 \hline
Field & Operator & Source\\
 \hline
 $g_{MN}$   &  $T_{\mu\nu}^\mathrm{em}$    & $\eta_{\mu\nu}$ \\ \hline
 $\lambda = e^{\phi}$ &   $\mathrm{tr}G_{\mu\nu}G^{\mu\nu}$  & $\lambda_t = g_\mt{YM}^2\Nc$ \\ \hline
 $T_{ij}$ &$\bar \psi^j \psi^i$ & $M_{ij}$ \\ \hline
 $A^{ij}_{M}$   & $\bar \psi^i \gamma_\mu \psi^j$ & $A_\mu^{({\rm ext})ij}$\\
 \hline
\end{tabular} 
\caption{The holographic dictionary. 
}
\label{tab:vqcd_dictionary}
\end{table}

We study the model in homogeneous, time-independent backgrounds, described by the five-dimensional metric with scale factor $A(r)$ and blackening factor $f(r)$,
\begin{equation} \label{metric}
\mathrm{d} s^{2}=e^{2A(r)}\left[-f(r) \mathrm{d} t^{2}+\mathrm{d} \mathbf{x}^{2}+\frac{\mathrm{d} r^{2}}{f(r)}\right]\ . 
\end{equation}
The boundary connecting to the field theory is located at a finite value of the holographic radial direction $r$ which we set to be zero, $r=0$. The action is formulated such that near the boundary the geometry is asymptotically AdS$_5$ and the blackening factor approaches one, $f(0)=1$. The function $A(r)$ is identified with the logarithm of the energy scale in field theory~\cite{Gursoy:2007cb,Gursoy:2007er}. In this background all bulk fields are functions of the holographic coordinate $r$ only.

The full action of the model reads,
\begin{equation}
\label{eq:actionfull}
    S_\mt{V-QCD} = S_\mt{g} + S_\mt{f} \,
\end{equation}
reflecting the two sectors of the model. The gluon sector is identified with the five-dimensional Einstein-dilaton gravity $S_\mt{g}$,
\begin{equation}\label{eq:action}
    S_\mt{g} = \Mp^3 \Nc^2 \int \dd^{5} x\sqrt{-\det g_{MN}}\left[R-\frac{4}{3 \lambda^2} g^{M N} \partial_{M} \lambda \partial_{N} \lambda+\Vg(\lambda)\right] +  2\Mp^3 \Nc^2 \int_\partial \dd^{4} x\sqrt{-h}\,{\mathcal{K}} \,
\end{equation}
which for an appropriate choice for the dilaton potential $\Vg$ and the Planck mass $\Mp$ matches the originally proposed IHQCD model~\cite{Gursoy:2009jd}. Here $\lambda(r) \equiv \exp(\phi(r)) $ is the exponential form of the dilaton field $\phi(r)$, which is identified with the 't Hooft coupling near the boundary~\cite{Gursoy:2007cb}, $R$ is the scalar curvature, and $h$ is the induced metric on the four-dimensional boundary manifold with $\mathcal{K}$ its extrinsic curvature. The second term in (\ref{eq:action}) is the standard Gibbons--Hawking term to render the variational problem well-defined.

The flavor sector~\cite{Bigazzi:2005md,Casero:2007ae} is inspired by the action of a pair of space filling tachyonic $D4$--$\overline{D4}$ branes. We only turned on the vectorial gauge fields, which merges the DBI actions of the $D4$ and the $\overline{D4}$ branes into a single DBI action simplifying to
\begin{align} \label{eq:Sfflavors}
S_\mt{f} & = -\Mp^3 \Nc \sum_{i=1}^{\Nf} \int \dd^5x\,\Vf(\lambda,\tau_i)\sqrt{-\det (g_{MN}  + w(\lambda,\tau_i) F_{MN}^{(i)}+\kappa(\lambda)\partial_M \tau_i\partial_N \tau_i)} \, \ .
\end{align}
Here $i=u,d,s$ represents the flavor dependence of the fields which are considered to be diagonal in flavor indices: the tachyon $T^{ij}=\tau_i\delta^{ij}$ and the gauge field $(A_M)^{ij} = A^{(i)}_M \delta^{ij}$.
$F^{(i)}_{MN}$ is the field strength tensor for the field $A^{(i)}_M$. This diagonal restriction for the tachyon at zero chemical potential is consistent with, and mandated by, the Vafa--Witten theorem~\cite{Vafa:1983tf,Cohen:2001hf}, which forbids the spontaneous breaking of vector-like flavor symmetries in QCD. Allowing off-diagonal tachyon components would generate flavor-violating condensates.
In the background, we only turn on the temporal components of the gauge fields $A^{(i)}_t$ as required for finite density configurations.  The flavor sector also contains Chern--Simons terms~\cite{Casero:2007ae,Jarvinen:2022mys,Raymond:2026twp} 
but they are irrelevant for the analysis in this article.
This flavor-dependent model~\cite{CruzRojas:2024etx,Jarvinen:2025mgj} was developed to describe QCD thermodynamics at finite temperature and baryon density, starting from the most general setup with arbitrary flavor-dependent quark masses and chemical potentials. Its fixed-point structure, relevant for zero- and finite-temperature RG flows, was also analyzed. For phenomenological applications, the model was then specialized to an effective \(2+1\)-flavor setup with two massless light quarks and one massive strange quark. We review here some main features of this phenomenological setup, see~\cite{Jarvinen:2025mgj} for more details.

To fully specify the model, one must determine the potentials $\Vg(\lambda)$, $\Vf(\lambda,\tau_i)$, $w(\lambda,\tau_i)$, and $\kappa(\lambda)$. These functions are fixed phenomenologically by comparison with QCD data, but their allowed form is strongly constrained by general properties of QCD. In particular, the qualitative agreement with features of QCD strongly constrains the asymptotic behavior of the potentials both at weak and strong coupling $\lambda$ as discussed in \cite{Gursoy:2007cb,Gursoy:2007er,Jarvinen:2011qe,Arean:2013tja,Jarvinen:2015ofa,Jarvinen:2021jbd}. 

Importantly, these qualitative requirements remain unaffected by the inclusion of the flavor dependence in the model. Therefore, we use the same basic Ans\"atze for the potentials as in the unflavored case has been used in \cite{Jokela:2018ers,Ishii:2019gta,Jokela:2020piw}, with minor modifications, following~\cite{Jarvinen:2025mgj} (see Appendix~\ref{app:st_potsform} for more details). The modifications include the addition of tachyon dependence $\tau_i$ in the potential $w$, taken in the form
\begin{align} \label{eq:wdef}
    w(\lambda, \tau_i) = \frac{w_\lambda(\lambda)}{(1+ \beta_s \tanh{(\gamma_s^2 \tau_i^2)})} \ ,
\end{align}
where $\beta_s$ and $\gamma_s$ are free parameters chosen to optimize the fit with lattice data at finite temperature and zero density, and $w_\lambda(\lambda)$ has the same functional form as in unflavored case, as summarized in the appendix \ref{app:st_potsform}.  In addition, the flavor potential $\Vf$
is modified as in \cite{Jarvinen:2022gcc,Jarvinen:2025mgj}, $\Vf(\lambda, \tau_i) = V_{\mt{f}\lambda} e^{-\tau_i^2} (1+ \tau_i^4)^{\tau_p}$
with the choice $\tau_p = 0.6 $ which makes the potential gentler than the case $\tau_p = 0$ and reduces the growth of the tachyon in the IR.
Together with the small adjustments of the parameters of $w$ and of the other DBI sector potentials, this makes it possible to reproduce the flavor dependence observed in lattice data.

Given the Ansatz and the potentials specified above, we can minimize the action to solve for the background geometries.
Since we are interested at finite temperature, the backgrounds can be divided into two classes of geometries: ``Thermal gas'' geometries with $f(r)=1$ and no horizon, and black hole backgrounds where $f(r)$ depends on the coordinate and there is a horizon, $f(\rh)=0$ at some finite value $r=\rh$. These black hole geometries are dual to deconfined phases and in this work we focus on these geometries. The black hole geometries further divide into two classes, which we call ``chirally symmetric'' and ``chirally broken'' black holes. This terminology refers to chiral symmetry in the light quark sector: as we ignore light quark masses, the chirally symmetric black holes have $\tau_u=0=\tau_d$ whereas these tachyon components are nonzero in the chirally broken geometries. Because we turn on a nonzero strange quark mass, chiral symmetry is always broken in the strange quark sector. As chiral symmetry is typically restored in deconfined phases, the chirally symmetric black holes are the most important solutions for the description of the deconfined phase. This was also seen in~\cite{Jarvinen:2025mgj} where chirally broken black holes were only found at temperatures much lower than the crossover temperature $\approx 150$~MeV. In this article, we only consider chirally symmetric black holes and leave the numerically much more demanding chirally broken solutions for future work. The solutions are constructed numerically following the recipe detailed in Appendix~\ref{app:details}. In particular, the nonzero value of the strange quark mass is obtained by requiring the strange quark tachyon component $\tau_s$ to have a nonzero source term (see Appendix~\ref{app:asymptotics}).

The model is tuned to lattice QCD thermodynamics in the deconfined phase at low baryon density, and is used to determine the EoS at high density and low temperature. As we pointed out in the introduction, doing so requires one to choose which data to fit. We will comment on this after setting up the holographic thermodynamics in Section~\ref{sec:thermo} (see also Appendices~\ref{app:cumulants} and~\ref{app:st_potsform}).

The phase-diagram analysis of \cite{Jarvinen:2025mgj} was not exhaustive. In particular, the chirally symmetric black hole sector was not systematically explored at finite density. It remained unclear whether this branch contains multiple competing solutions, whether phase transitions occur within it, and how such structure affects the dense-matter EoS. The goal here is to address these questions by performing a systematic scan of chirally symmetric black hole solutions over the relevant parameter space and by computing the corresponding EoS.

\section{Setting up the thermodynamics} \label{sec:thermo}
In this section, we discuss the setup for the holographic thermodynamics and how this is taken into account when comparing the model to data. We first summarize the holographic dictionary and the thermodynamic relations used to extract the relevant observables from the bulk solution. 

The temperature and entropy are given by the surface gravity and the area of the black hole, yielding
\begin{equation} \label{eq:BHthermo}
    T = \frac{1}{4\pi}\left|f'(\rh)\right|\ , \qquad s = 4\pi \Mp^3 \Nc^2 e^{3 A(\rh)} \ , 
\end{equation}
respectively.
The chemical potentials are defined as the boundary values of the temporal components of the gauge fields, $\mu_i = A_t^{(i)}(0)$, where $i$ runs over different flavors, $i=u,d,s$. The temporal component of the gauge fields can be integrated out, leading to~\cite{Jarvinen:2025mgj}
\be\label{eq:holochem}
\mu_i  =  \bar n_i\int_0^{\rh} \dd r \frac{e^{-A}\sqrt{1+e^{-2A}f \kappa(\lambda)\dot \tau_i^2}}{\Vf(\lambda,\tau_i)w(\lambda, \tau_i)^2\sqrt{1+ \frac{ e^{-6 A}  \bar  n_i^2}{ w(\lambda,\tau_i)^2 \Vf(\lambda,\tau_i)^2}}} \ ,
\ee
where the integration constants $\bar n_i$ are related to the physical quark densities $n_i$ by
\be
 n_i = \Mp^3 \Nc \bar n_i \ . 
\ee

As per the holographic dictionary, the internal energy density $\epsilon$ and the pressure $P$ satisfy the Smarr relation and the first law of thermodynamics: 
\begin{align}\label{eq:epsplusP}
\epsilon + P &= Ts + n_u\mu_u+n_d \mu_d+n_s\mu_s \ ;   \\
\dd P &= s\dd T+n_u \dd\mu_u+n_d \dd\mu_d+n_s \dd \mu_s \ . 
\label{eq:firstlaw}
\end{align}
The pressure and energy density can in principle be computed directly by analyzing the on-shell action and the near-boundary asymptotics of the geometry, but in V-QCD it is simpler to use the relations~\eqref{eq:firstlaw} and~\eqref{eq:epsplusP}.

The system has three conserved charges. It is convenient to define them in the basis taking into account the coupling to the electroweak sector, so that the conserved charges are the baryon number B, charge Q, and strangeness S. The relationships between the quark and conserved charge bases are
\begin{align}
\nB &= \frac{1}{3}(n_u+n_d+n_s) \label{eq:nB} \\
\nQ &= \frac{2}{3}n_u -\frac{1}{3}n_d-\frac{1}{3}n_s \label{eq:nQ}  \\ 
    n_\mathrm{S} &= -n_s \label{eq:nS} \ 
\end{align}
between the densities and 
\begin{align}
\label{eq:muu}
\mu_u& = \frac{1}{3}\muB+\frac{2}{3}\muQ\\
\mu_d& = \frac{1}{3}\muB-\frac{1}{3}\muQ \label{eq:mud} \\
\mu_s&=\frac{1}{3}\muB-\frac{1}{3}\muQ-\muS  \label{eq:mus}
\end{align}
between the chemical potentials. 
In the conserved charge basis~\eqref{eq:firstlaw} becomes simply
\begin{equation}
    \dd P = s\dd T + \nB \dd \muB+\nQ\dd\muQ+n_\mathrm{S}\dd\muS \ .
\end{equation}

The full phase diagram of the model therefore depends on four thermodynamical parameters: the temperature and three chemical potentials. We will not attempt to explore the full parameter space, but instead choose conditions that restrict the phase diagram to depend on two parameters, chosen to be the temperature $T$ and the baryon number chemical potential $\muB$. To this end, as explained in the introduction, we will be interested in two settings: $\beta$-equilibrium and heavy-ion settings. They are defined as follows:
\begin{enumerate}
    \item \textbf{$\beta$-equilibrium.} As the first setting, in part motivated by matter in neutron stars, we consider charge neutral quark-gluon plasma in $\beta$-equilibrium. For charge neutrality we require $\nQ =0$ and for $\beta$-equilibrium we take $\mu_s = \mu_d$. These correspond to charge neutral and $\beta$-equilibrated matter in the absence of leptons.\footnote{This configuration is also close to $\beta$-equilibrium if one adds the leptons, since their number densities are typically small.} The charge neutrality requirement translates to $n_u = (n_d+n_s)/2$ in terms of quark densities, while the $\beta$-equilibrium condition for the down-type quarks can also be written as $\muS=0$ as seen from~\eqref{eq:mud} and~\eqref{eq:mus}. Note that these conditions imply that $\mu_u \ne \mu_d$ (except when all chemical potentials vanish). This setup we call $\beta$-equilibrated quark-gluon plasma in the following.
    \item \textbf{Heavy-ion collisions.} As the second setting, we choose constraints that mimic the conditions in the quark-gluon plasma created in heavy-ion collisions. Although the formation of a quark-gluon plasma strongly enhances the production of strange quarks~\cite{Rafelski:1982pu,ALICE:2016fzo}, the strong interaction produces strange and anti-strange quarks in exactly equal numbers. Because the initial colliding nuclei carry zero net strangeness and weak decays are far too slow to occur during the lifetime of the fireball, net flavor is conserved, enforcing the constraint $\nS=0$. Neglecting leptons and mimicking the composition of the nuclei we fix the ratio of charge to baryon density to $\nQ/\nB = 0.4$.  This setup will be called HIC in the text in reference to heavy-ion collisions.
\end{enumerate}

\begin{figure}[!htb]
\centering
    \includegraphics[width=0.7\textwidth]{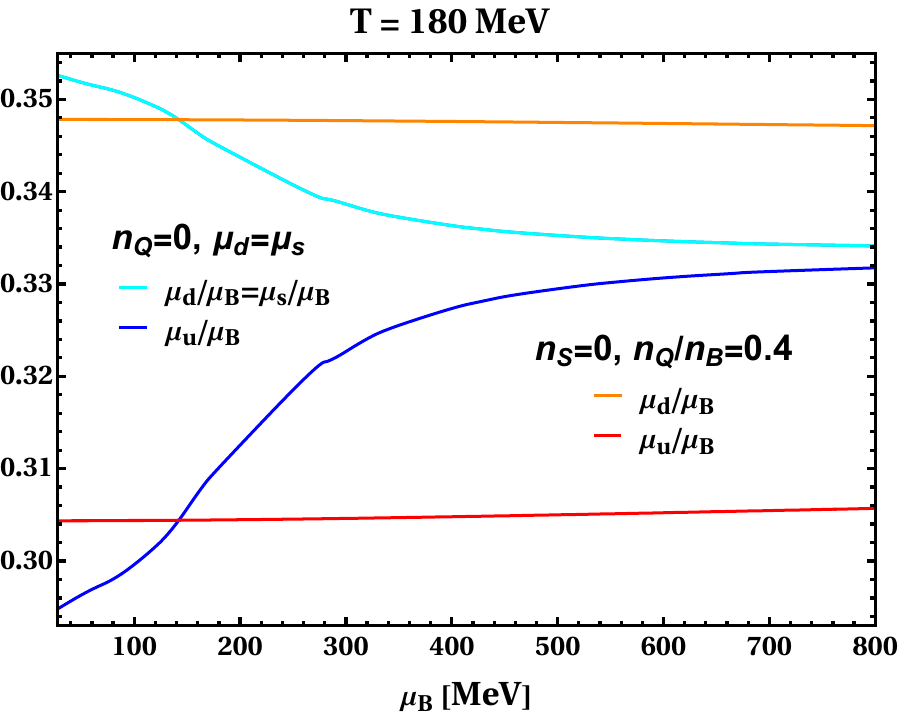}
     \caption{Evolution of the scaled quark chemical potentials $\mu_i/\muB$ as a function of the baryon chemical potential $\muB$ at $T=180$ MeV for the $\beta$-equilibrium (bendy curves) and HIC (slowly evolving curves) setups in V-QCD. Because the $\beta$-equilibrium conditions enforce $\mu_d=\mu_s$, these identical chemical potentials are shown as a single curve (cyan). Conversely, in the HIC setup, the $\mu_s$ curve is omitted entirely, as the constraint of vanishing net strangeness $\nS=0$ dictates that $\mu_s=0$ directly via (\ref{eq:holochem}) and~\eqref{eq:nS}.}
     \label{fig:chems}
\end{figure}

To explicitly demonstrate how these distinct environmental constraints dictate the composition of the deconfined quark matter, Fig.~\ref{fig:chems} illustrates the behavior of the individual quark chemical potentials, scaled by the baryon chemical potential, $\mu_i/\muB$, as a function of $\muB$ at a representative temperature of $T=180$ MeV. In the $\beta$-equilibrated setting, the weak interaction equilibrium enforces $\mu_d=\mu_s$, tying the down and strange quark dynamics closely together across all densities. Conversely, in the HIC setting, the strict strangeness neutrality condition $\nS=0$ forces the strange quark chemical potential to vanish entirely $\mu_s=0$. As a result, the thermodynamic states evaluated in these two settings sample completely different regions of the full multidimensional chemical potential space, which ultimately drives the drastic differences in their resulting phase structures to which we continue next.  

Let us then discuss how the thermodynamic formulas become in these two cases and how phase equilibrium can be determined. In the case of $\beta$-equilibrium, starting from the thermodynamics in terms of the conserved quantities, we impose the conditions
\be \label{eq:betaeqconds}
 \muS = 0\ , \qquad \nQ=0 \ .
\ee  
The charge condition suggests that we should carry out a Legendre transformation from $\muQ$ to the conjugate variable $\nQ$. However, since we set the charge to zero the transformation term $\nQ \muQ$ vanishes. Therefore the correct equilibrium condition is simply the mechanical, thermal and chemical equilibrium:
\begin{equation}
P^\text{(1)} = P^\text{(2)}\ ,  \quad  T^{(1)} = T^\text{(2)}\ , \quad \muB^\text{(1)} = \muB^\text{(2)} \ ,
\end{equation}
where the indices denote the quantities in two different phases, 1 and 2. Recall that in addition to these conditions, we require~\eqref{eq:betaeqconds} in both phases. Therefore, while the baryon and strangeness chemical potentials are matched at the transition, the charge chemical potential is not: $\muQ^\text{(1)} \ne \muQ^\text{(2)}$ in general. This means that the transition is a forced congruent transition (see, \eg,~\cite{Iosilevskiy:2010qr,Hempel:2013tfa}). 
If we allow the two phases to independently rearrange their composition near an interface, a mixed phase appears, where $\nQ\ne 0$. This arrangement is often called the ``Gibbs construction.''  We will not consider such constructions in this article. 
The first law~\eqref{eq:firstlaw} constrained for $\beta$-equilibrium conditions can be written as 
\begin{equation} \label{eq:dPbetaeq}
\dd P = s \dd T + \nB \dd\muB\ .
\end{equation}

The situation in the heavy-ion case is more complicated (see, \eg,~\cite{Hempel:2013tfa}). In terms of the conserved charges, the conditions can be written as
\be
 n_\mathrm{S}=0\ , \qquad \tilde n \equiv \nQ - \alpha \nB = 0 \ ,
\ee  
where $\alpha \approx 0.4$ is fixed. We would like to impose these conditions and draw the phase diagram as a function of $T$ and $\muB$. However, this does not work simply, due to the condition $\tilde n = 0$, because the transformation from $(\muB,\muQ)$ to $(\muB,\tilde n)$ is not canonical. Instead, the transformation to $(\tilde \mu, \tilde n)$ is canonical, where 
\begin{equation} \label{eq:mueff}
\tilde{\mu} = \muB+ \alpha \muQ \ .
\end{equation}
That is, the constrained ensemble as a function of $T$ and $\tilde \mu$ is well defined. As in the $\beta$-equilibrium case there is no need to carry out Legendre transformation explicitly, and the equilibrium conditions are given by
\begin{equation}
P^\text{(1)} = P^\text{(2)}\ ,  \quad  T^{(1)} = T^\text{(2)}\ , \quad \tilde \mu^\text{(1)} = \tilde \mu^\text{(2)} \ .
\end{equation}
Similar to the $\beta$-equilibrium case, this is a congruent transition. The two coexisting phases are not allowed to independently rearrange their composition. Instead, the conditions $n_\mathrm{S}=0$ and $n_\mathrm{Q}/n_\mathrm{B}=\alpha$ are imposed separately in each phase, forcing both phases to have the same local composition. Only a single chemical potential, $\tilde \mu$, is matched at the transition in this case. The pressure differential can be written as
\begin{equation}\label{eq:dpqgp}
    \dd P =  s \dd T + n_u\left(\dd\mu_u+\frac{2-\alpha}{1+\alpha} \dd\mu_d \right) = s\dd T + \nB\dd\tilde \mu \ .
\end{equation}

The differentials~\eqref{eq:dPbetaeq} and~\eqref{eq:dpqgp} will then be used to compute the pressure numerically by integrating them over paths in the parameter space (see Appendix~\ref{app:details} for details). In principle, the pressure can also be computed by evaluating the on-shell action or by studying the near-boundary asymptotics of the geometry. However, while this is possible in our model, it is technically highly demanding due to the complex near-boundary behavior of the geometry~\cite{Jarvinen:2011qe}. Therefore, it is simplest to carry out the integration of the pressure by using the first law. In this integration, as in~\cite{Jarvinen:2025mgj}, the integration constant is fixed by requiring the pressure of the confined vacuum to vanish.

Let us then discuss the various susceptibilities. The most important ones, both when comparing to the lattice results and to data from heavy-ion collisions, are the  dimensionless\footnote{Note that all susceptibilities discussed in this article are dimensionless. Therefore, for notational simplicity, we do not follow the common notation where dimensionless susceptibilities have hats.} baryon number susceptibilities 
\begin{equation}\label{eq:chindefunconst}
\chi^\mathrm{B}_n(T,\muB,\muQ,\muS)
= T^{n-4}\frac{\partial^n P(T,\muB,\muQ,\muS)}{\partial\muB^n} 
\end{equation}
The judicious choice of the temperature prefactor compensates for the canonical dimensions of the pressure and chemical potential, yielding a dimensionless quantity.
On lattice one is often interested in these  susceptibilities  at zero chemical potential $\muB = \muQ = \mu_\mathrm{S} = 0$, whereas the heavy-ion data probes them at nonzero chemical potential.

The flavor dependent configurations we consider here (both $\beta$-equilibrium and HIC) fix the strangeness and charge chemical potentials in terms of the baryon chemical potential. This results in 
\be
P(T,\muB,\muQ(\muB),\muS(\muB)) = P_\text{eff}(T,\muB)\ .
\ee
The constrained susceptibilities are defined as the derivatives of the pressure with respect to the baryon chemical potential
\be \label{eq:chindef}
 \chinT(T,\muB) = T^{n-4}\frac{\partial^{n}P_\text{eff}(T,\muB)}{\partial\muB^{n}} \ . 
\ee

The second-order quark susceptibilities at zero density are defined as
\begin{equation}
\chi^{ij}(T) = \frac{1}{T^2}\frac{\partial^{2} P(T,\mu_u,\mu_d,\mu_s)}{\partial\mu_i\partial\mu_j}\bigg{\lvert}_{\mu_k=0} 
\end{equation}
with $i,j,k\in \{u,d,s\}$.  
When the baryon susceptibilities are expressed in terms of the quark susceptibilities we in general have contribution also from the off-diagonal quark susceptibilities. This is already evident for the second-order baryon susceptibility,
\begin{equation} \label{eq:chi2Btransf}
\chi^\mathrm{B}_2 = \frac{1}{9}\left( \chi^{uu}+\chi^{dd}+\chi^{ss} + 2\chi^{ud}+2\chi^{us}+2\chi^{ds}\right) = \frac{1}{9}\left( 2\chi^{ll}+\chi^{ss} + 2\chi^{ud}+4\chi^{ls}\right) \ ,
\end{equation}
where the last expression is valid for identical light quarks, with $\chi^{uu}=\chi^{dd}\equiv \chi^{ll}$ and $\chi^{us}=\chi^{ds}\equiv \chi^{ls}$. This formula therefore highlights the issue discussed above in the introduction: according to lattice data~\cite{Borsanyi:2011sw} $\chi^{ls}$ is nonzero near the crossover, and due to the large numerical coefficient in~\eqref{eq:chi2Btransf}, this affects rather strongly the baryon number susceptibility in this region. Because the holographic model predicts that all off-diagonal susceptibilities vanish at zero density, we have two functions $\chi^{ll}$, $\chi^{ss}$ from the model to fit three independent variables in lattice data, so one needs to choose how the fit is done. The symmetric choice in~\cite{Jarvinen:2025mgj}, which we follow here, is to fit the data for $\chi^{ll}$ and $\chi^{ss}$, and ignore the mismatch of $\chi^{ls}$ (or equivalently, ignore $\chi^\text{B}_2$). See Appendix~\ref{app:cumulants} for the fit.

\section{Fate of the critical point in the purely holographic model}\label{sec:betaequil}

In this section, we discuss results from the ``purely holographic'' V-QCD model, \ie, predictions directly from the holographic model without trying to incorporate the physics of the meson gas
at low temperature and density. We focus on the critical point and its signatures in the predictions for various cumulants that can be measured at RHIC.  To construct a globally consistent phase diagram, we will complement these holographic results in Section~\ref{sec:heavyion} by incorporating a low-temperature hadronic sector using HRG models.

Before discussing the thermodynamic results, we must clearly define the scope of the holographic phase diagrams presented in this section.  Here, we evaluate the phase structure entirely within the purely holographic model of the chirally symmetric black hole branch of V-QCD. We omit the confined holographic nuclear matter phase. This phase is expected to become important at high density, in the regime relevant for neutron star physics, which is beyond the scope of the study in this article. Furthermore, as we explained in Section~\ref{sec:setup}, we also ignore the chirally broken deconfined phase because the numerical construction of this phase is challenging.\footnote{To be precise, chiral symmetry is explicitly broken in the strange sector via the s quark mass source. However, the u and d quarks remain massless, so chiral symmetry is not explicitly broken in the light sector. By the ``chirally broken deconfined phase'' we refer here to spontaneous condensation of these light quarks.} According to the zero density results of~\cite{Jarvinen:2025mgj}, this phase appears only at temperatures much lower than the crossover temperature in QCD, \ie, outside the region of focus of this study at least at small densities. Moreover, we do not turn on explicit mass sources for the u and d quarks, nor do we track their corresponding condensates. Our primary objective in this section is to study the properties of the deconfined medium and the CEP. That is, the transition lines discussed below represent the thermodynamics of the deconfined quark-gluon plasma. 

Following the above discussion, we restrict our focus on the chirally symmetric sector, using the holographic action matched to $\Nf=2+1$ flavor zero density lattice data. As we shall show, this leads to a critical point on the phase diagram. This critical point appears in the deconfined phase, and is the endpoint of a first-order transition within this phase, holographically implemented as a transition between two black hole phases. The flavor-dependent V-QCD framework then allows us to  test explicitly how different physical environments (such as those defined in Section~\ref{sec:thermo}) affect the critical point.

\subsection{The influence of environmental conditions on the phase diagram}\label{sec:envi}

\begin{figure}[!htb]
    \includegraphics[width=0.5\textwidth]{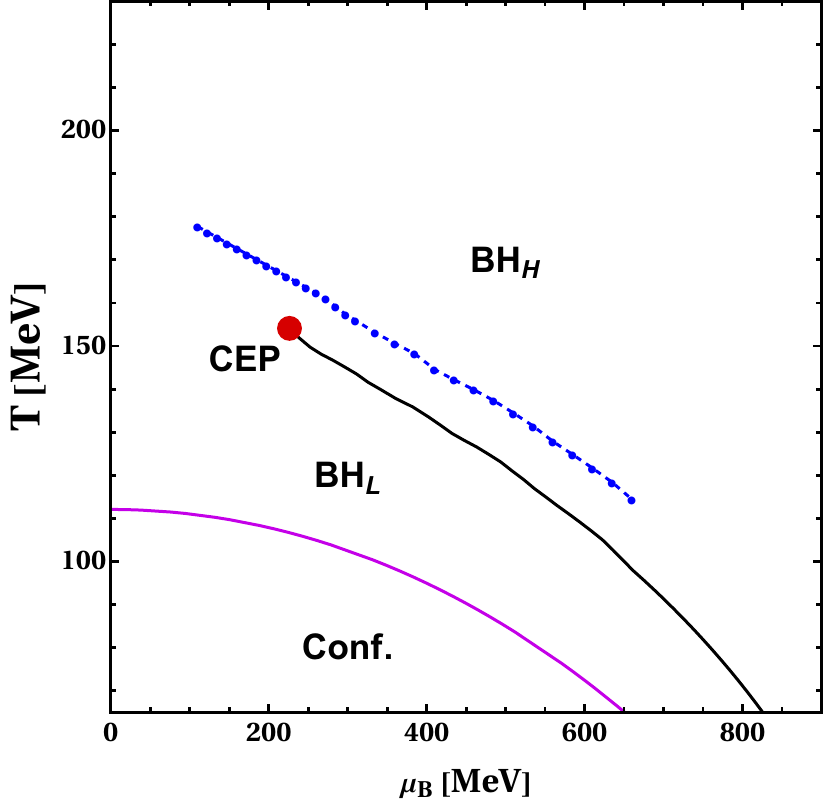}
    \includegraphics[width=0.5\textwidth]{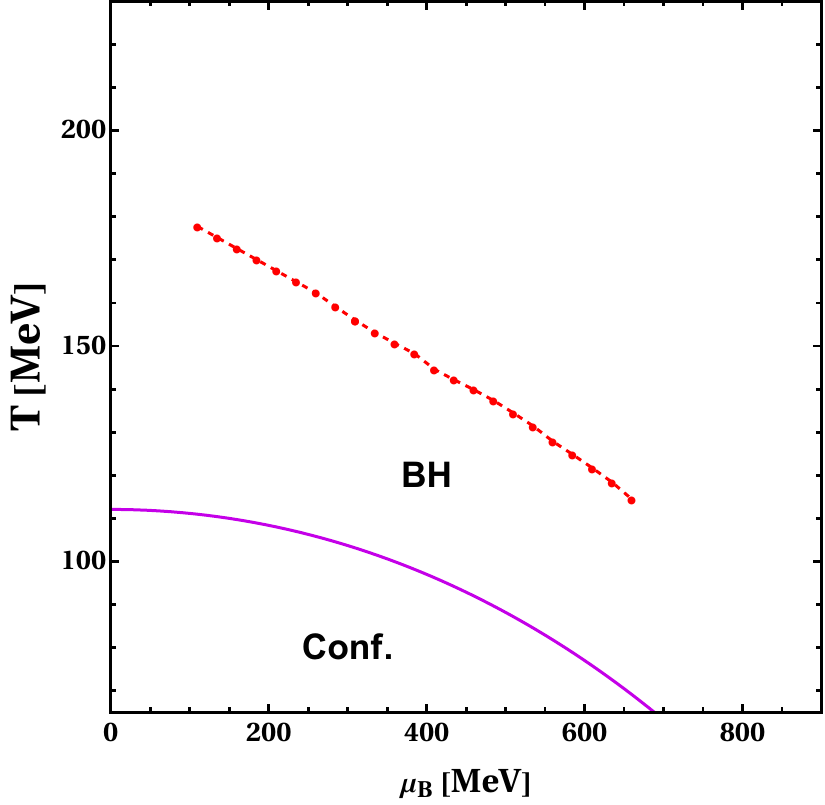}
     \caption{{\bf{Left:}} The purely holographic phase diagram in the ($\muB$,$T$)-plane for the V-QCD model under charge neutrality and $\beta$-equilibrium. The black line (FOPT) shows a first-order phase transition between two black hole phases ($\text{BH}_L$ and $\text{BH}_H$), terminating at a critical endpoint (CEP).  {\bf{Right:}} The corresponding phase diagram for the V-QCD model under strangeness neutrality and $\nQ/\nB=0.4$ (HIC setup). The FOPT and CEP are absent in this panel. In both plots, the purple line corresponds to the confinement transition where the pressure of the black hole solution goes negative, and the blue (red) dots connected by dashed lines show where the cumulant ratios of Fig.~\ref{fig:C4C2step} are evaluated in the $\beta$-equilibrium (HIC) scenario.}
     \label{fig:holophasediagram}
\end{figure}

In Fig.~\ref{fig:holophasediagram} we present the resulting purely holographic V-QCD phase diagrams in the $(\muB,T)$-plane, demonstrating the striking difference between the two dense-matter environments defined in Section~\ref{sec:thermo},  \ie, the $\beta$-equilibrium (left panel) and HIC (right panel). 

Interestingly, 
in the $\beta$-equilibrated environment of the left panel, the model exhibits a prominent first-order phase transition between two distinct deconfined states, denoted by the black line terminating at the CEP. The location of the CEP is approximately $(\muB,T) \approx (225,155)$~MeV. The two black hole phases are labeled BH${}_\text{H}$ (high temperature/density) and BH${}_\text{L}$ (low temperature/density). Moreover, the purple line indicates the confinement transition, below which the pressure of the deconfined state ${\sim\cal{O}}(\Nc^2)$ corresponding to the black hole solution becomes negative, and a confining geometry with parametrically small ${\cal{O}}(\Nc^0)$ pressure dominates.

Conversely, Fig.~\ref{fig:holophasediagram} (right), display the exact same V-QCD model evaluated under the relativistic heavy-ion collision conditions, enforcing strict strangeness neutrality and a fixed charge-to-baryon ratio. While the confinement transition (purple line) remains, the deconfined FOPT and its associated CEP completely vanish. The environmental conditions have therefore a strong effect on the thermodynamic landscape.

These findings offer a compelling interpretation of 
recent experimental data~\cite{STAR:2025ydk}. High-statistics measurements from the STAR collaboration's BES-II program have reported no evidence of a critical point or a first-order phase transition. While it is tempting to conclude that no CEP exists anywhere in the QCD phase diagram, our results demonstrate that a FOPT can easily exist in the underlying theory, but remain entirely hidden by the specific environmental conditions of the heavy-ion setup.

The result for the critical point in $\beta$-equilibrium is strikingly different from earlier studies of the V-QCD phase diagram where a version of the model was used that was compared to lattice data but did not include flavor dependence~\cite{Jokela:2018ers,Jokela:2020piw,CruzRojas:2024igr}: in these earlier works, no such CEP was reported. Actually, to be precise, a similar CEP between black hole phases does exist in these works, but it is located in the region of negative pressure, which is subdominant to the confined phase represented by a horizonless geometry. The location of the critical point in Fig.~\ref{fig:holophasediagram} (left) is also different from what is obtained in EMD~\cite{Jokela:2024xgz,Li:2025lmp} and DBI~\cite{Jokela:2024xgz} fits following the approach of~\cite{DeWolfe:2010he} as it appears at much lower chemical potential than in these works. Moreover, it is similarly located at much lower density than the critical point obtained for the unflavored V-QCD setup after comparing with HRG models~\cite{Demircik:2021zll,Ecker:2025vnb}. We will comment on this more in Section~\ref{sec:heavyion} where we compare the current flavored setup to HRG models. As far as we can see these differences are not solely a direct effect of the addition of the strange quark mass, but arise because the strategy of fitting the data (motivated by the addition of the quark flavors), is different: as explained above, we fit the diagonal components of the quark susceptibility matrix while ignoring the off-diagonal entries.

We also wish to study how chiral symmetry behaves across the FOPT of Fig.~\ref{fig:holophasediagram}  (left). However, extracting the value of the chiral condensate in the V-QCD model at finite quark mass is technically involved~\cite{Jarvinen:2015ofa}. Therefore instead of the condensate $\langle \bar ss\rangle$ of the strange quarks, we analyze the horizon value $\tau_{h}$ of the strange quark component of the tachyon field. The near-boundary behavior of the tachyon is schematically $\tau_s \sim m_s r + \langle \bar ss\rangle r^3$, where $m_s$ is the strange quark mass (see Appendix~\ref{app:asymptotics}). Because the mass is relatively low, the source term is rather small, so that the horizon value of the solution reflects the vacuum expectation value term, and therefore acts as a proxy for spontaneous component of chiral symmetry breaking in the strange quark sector.

\begin{figure}[!htb]
    \includegraphics[width=0.5\textwidth]{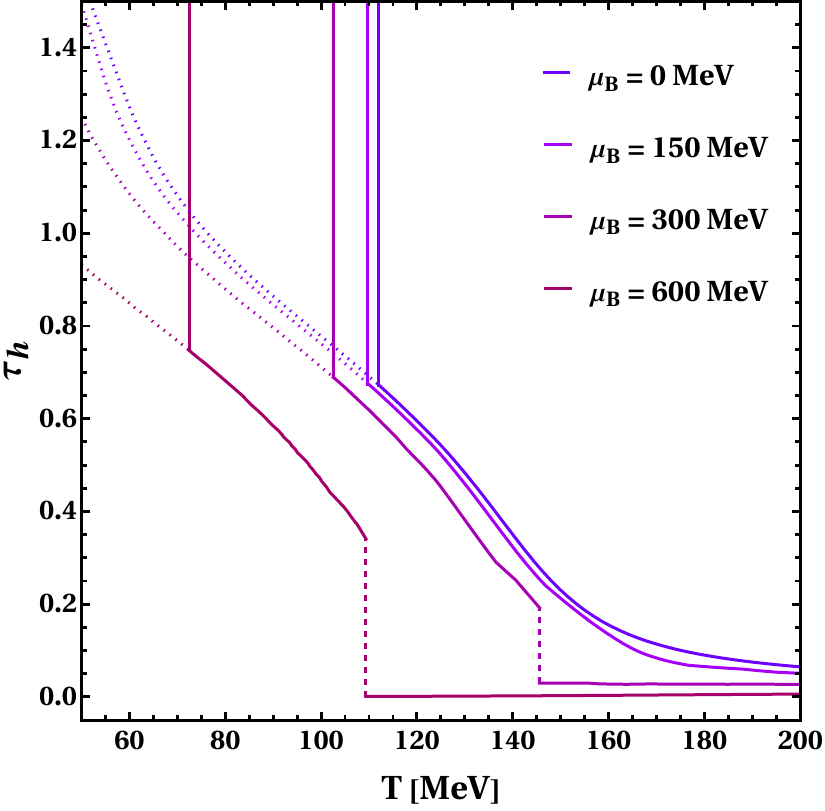}
    \includegraphics[width=0.5\textwidth]{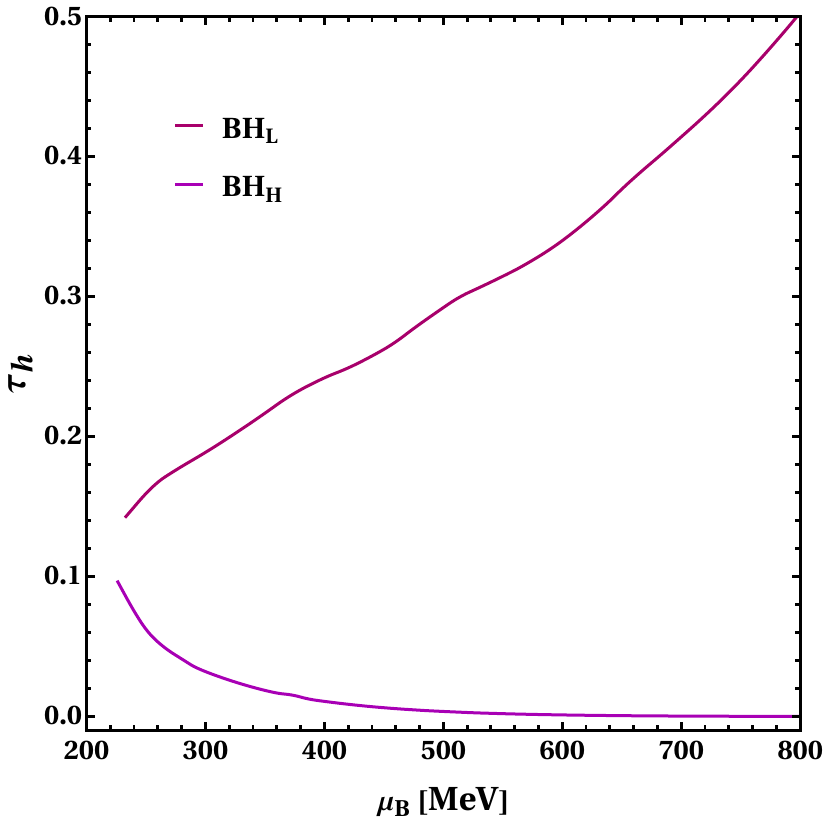}
     \caption{{\bf{Left:}} The horizon value of the tachyon as a function of the temperature is shown at four different baryon chemical potentials. In the confining phase,  shown in the left panel of Fig.~\ref{fig:holophasediagram}, the tachyon diverges at the horizon as indicated by the vertical solid lines. The dotted portions of the curves correspond to the result in the subdominant low-temperature phase in this region. The vertical dashed lines indicate the discontinuity of the condensate at the BH${}_\text{L}$ to BH${}_\text{H}$ transition. {\bf{Right:}} The horizon value of the tachyon as a function of baryon chemical potential is shown across the black-hole transition of Fig.~\ref{fig:holophasediagram}.}
     \label{fig:tau}
\end{figure}

We find that the FOPT of Fig.~\ref{fig:holophasediagram}  (left) is characterized by a discontinuous jump in the horizon value of the tachyon field (see Fig.~\ref{fig:tau}). In the BH${}_\text{L}$ phase, the tachyon retains a non-zero value, but as the system crosses into the BH${}_\text{H}$, it drops to approximately zero. Thus, the natural physical interpretation of this boundary is a chiral restoration transition; even though chiral symmetry is explicitly broken by the massive strange quark, the effective mass becomes parametrically zero in the high-density BH${}_\text{H}$ phase. This is illustrated in the left panel of Fig.~\ref{fig:tau} where the horizon value for the tachyon is shown at four different baryon chemical potentials. On the right panel the tachyon horizon value is evaluated across the black-hole transition, \ie, in both phases at the transition line. We see that the values converge towards the critical point\footnote{The two curves of the right panel meet at the phase transition, but this is not seen in the figure due to the rapid variation of the horizon values and the limited resolution of our numerical grid. The curves in the immediate vicinity of the CEP are not reliable.} but for bulk of the transition, in BH${}_\text{H}$ phase the horizon value is practically zero. 

Finally, it is worth noting that various non-holographic approaches, such as fRG studies~\cite{Fu:2019hdw}, Dyson--Schwinger equations~\cite{Fischer:2018sdj,Gunkel:2021oya}, effective chiral models~\cite{Fukushima:2008wg}, and lattice QCD extrapolations via Pad\'e approximants~\cite{Basar:2023nkp}, or constant entropy contours~\cite{Shah:2024img,Shah:2026mzu}, do predict a critical point at finite density. Although recent lattice-based work has extended the constant entropy scheme to map a critical surface across the full chemical potential space~\cite{Shah:2026mzu}, these extrapolations inherently rely on low-order $\mathcal{O}(\mu^2)$ expansions anchored at $\mu=0$. More broadly, non-holographic models typically evaluate the thermodynamics along simplified trajectories, such as the $\muS=0=\muQ$ axis. As our holographic results demonstrate, moving beyond low-density expansions to strictly enforce the heavy-ion density conditions ($\nS=0$ and $\nQ/\nB=0.4$) qualitatively modifies the resulting phase structure. This highlights a compelling opportunity for future theoretical work: explicitly implementing these stringent conditions within fRG or effective model computations would provide a rigorous test of whether the vanishing of the critical point is a universal feature of heavy-ion collisions. See~\cite{Wen:2019ruz,Gunkel:2021oya,Fu:2026qnl} for work in this direction.

\subsection{Net-proton fluctuations at chemical freeze-out}\label{sec:HRGvsVQCDcumulants}

\begin{figure}[!htb]
\centering
    \includegraphics[width=0.7\textwidth]{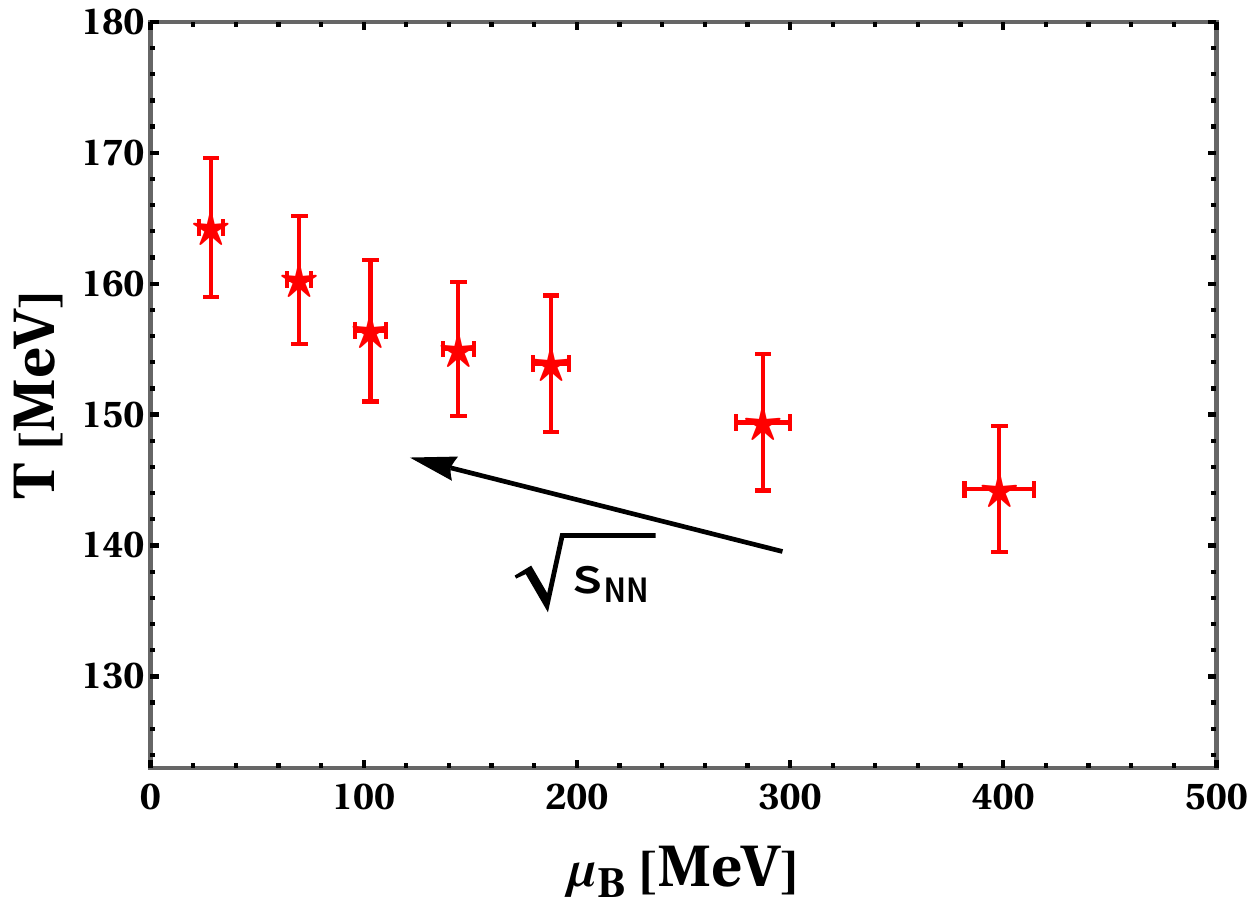}
    \caption{The chemical freeze-out parameters for the collision events $\sqrt{s_\mathrm{NN}} \in \{7.7,11.5,19.6,27,39,62.4,200\}$ GeV in the $(\muB,T)$-plane, represented by the red stars and their corresponding uncertainty. The arrow shows the direction of increasing $\sqrt{s_\mathrm{NN}}$. The data is taken from Table VIII of \cite{STAR:2017sal} for 0-5~\% centrality.}
    \label{fig:diagram}
\end{figure}

We now compare our holographic results with experimental data from heavy-ion collisions. In high-energy heavy-ion collisions, the key controllable variables are the colliding ion species, the collision centrality, and the center-of-mass collision energy, $\sqrt{s_\mathrm{NN}}$. Experimental observables are typically reported as a function of $\sqrt{s_\mathrm{NN}}$. In contrast, our holographic model characterizes the thermodynamic state using the temperature and baryon chemical potential. Therefore, we must map a given coordinate in the $(\muB,T)$-plane to the corresponding collision energy. Collision data allows for this mapping at the chemical freeze-out boundary. We adopt the empirical freeze-out parameters reported in Table VIII of \cite{STAR:2017sal} for 0 to 5~\% centrality. These parameters are illustrated in Fig.~\ref{fig:diagram}. The errors shown in $T$ and $\muB$ directions are incorporated into the V-QCD results using first-order error propagation discussed in Appendix~\ref{app:pandsusc}.
 
Next, we compare our holographic predictions with the BES data from the STAR collaboration \cite{STAR:2021iop}. We focus on the higher moments of the distributions of conserved quantities, $(N)$, because they are highly sensitive to the correlation length~\cite{Stephanov:2011pb,STAR:2020tga}. While these moments can be analyzed for any conserved charge, we restrict our focus to the net-proton number measured in the collisions. Of particular relevance are the skewness $S=\langle (\delta N)^3\rangle/\sigma^3$ and the kurtosis $\kappa=[\langle (\delta N)^4\rangle/\sigma^4]-3$ of the net-proton number fluctuations $\delta N=N-M$,  where $M=\langle N\rangle$ is the mean and $\sigma$ is the standard deviation. To translate between the thermodynamic susceptibilities $\chi^\mathrm{B}_n$ defined in (\ref{eq:chindefunconst}), and the experimental net-proton cumulants $C_n$ from \cite{STAR:2021iop}, where $C_1=M$, $C_2=\sigma^2$, $C_3=S\sigma^3$, and $C_4=\kappa\sigma^4$, we use the standard volume relation~\cite{Gupta:2011wh} 
\be\label{eq:hiccum}
C_n = V_\mathrm{fr}T^3\chi^\mathrm{B}_n \ ,
\ee
where $V_\mathrm{fr}$ is the average freeze-out volume of the quark-gluon plasma produced in the collision. 
Because the definition of this volume introduces significant uncertainties, we only evaluate the ratios between different cumulants, ensuring that the volume factor cancels.\footnote{The formula \eqref{eq:hiccum} also receives corrections due to finite fireball size, which we ignore here~\cite{Vovchenko:2020gne}. Such corrections  only partially cancel in the ratios.}
The experimental data are provided as ratios, specifically $C_2/C_1=\chi^\mathrm{B}_2/\chi^\mathrm{B}_1$, $C_3/C_2=S\sigma=\chi^\mathrm{B}_3/\chi^\mathrm{B}_2$, and $C_4/C_2=\kappa\sigma^2=\chi^\mathrm{B}_4/\chi^\mathrm{B}_2$. Note that $\chi^\mathrm{B}_1 = \nB/T^3$ and $M = V_\mathrm{fr} \nB$. 
Further details on the evaluation of the cumulants are provided in Appendix~\ref{app:pandsusc}. 

\begin{figure}[!htb]
    \centering
    \includegraphics[width=0.49\textwidth]{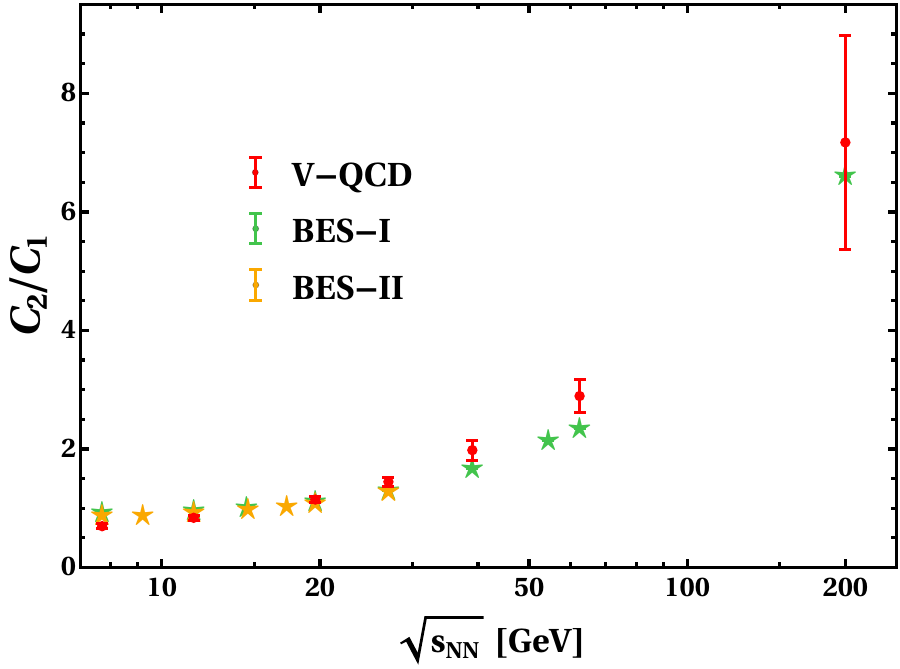}
    \includegraphics[width=0.49\textwidth]{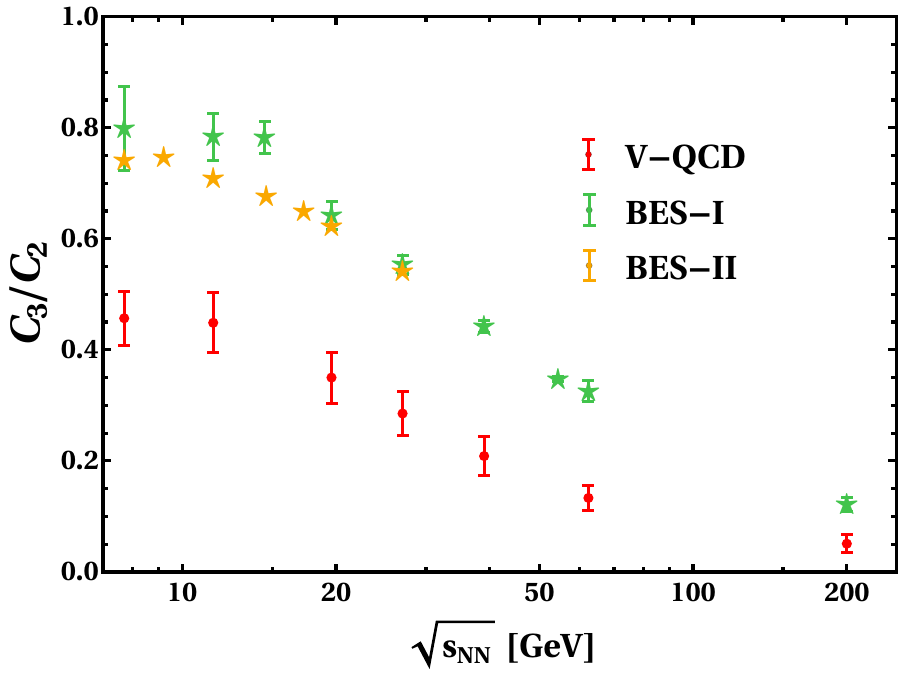}

    \vspace{0.5cm}
    \includegraphics[width=0.5\textwidth]{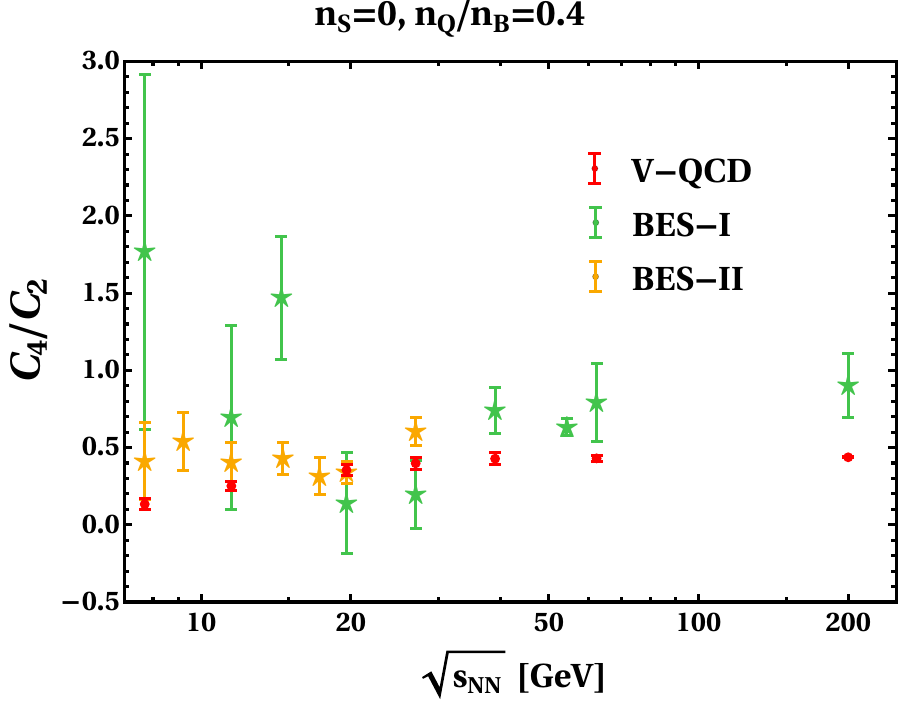}

    \caption{Ratios of cumulants as a function of the collision energy $\sqrt{s_\mathrm{NN}}$. The red results are computed from the V-QCD model with HIC conditions. The green BES-I collision data is from \cite{STAR:2021iop} for Au+Au 0–5\% centrality. The plotted collision energies are $\sqrt{s_\mathrm{NN}}\in\{7.7,11.5,14.5,19.6,27,39,54.4,62.4,200\}$ GeV. We also include preliminary and published data from BES-II~\cite{Pandav:CPOD2024,STAR:2025zdq} for comparison, marked by the orange stars. The uncertainty in the mapping 
    $\sqrt{s_\mathrm{NN}} \leftrightarrow (\muB,T)$ gives the errors shown in the V-QCD results.}
    \label{fig:qgpcumulants}
\end{figure}

Our results for the cumulant ratios in the HIC setup are presented in Fig.~\ref{fig:qgpcumulants}. The variance-to-mean ratio $C_2/C_1$ matches the  heavy-ion data well, but the model underestimates the ratio $C_3/C_2$ by a factor which is roughly constant. This discrepancy originates from our fit to lattice data; the discrepancy and a partial remedy are further discussed in Appendix~\ref{app:cumulants}. Nevertheless, the qualitative behavior is clear: the higher cumulants show no sign alternation and lack the pronounced non-monotonic dips with respect to the collision energy $\sqrt{s_\text{NN}}$. There is no indication of critical fluctuations, which is entirely consistent with our framework, since the FOPT and CEP are structurally absent in the HIC setup as demonstrated earlier in Fig.~\ref{fig:holophasediagram} (right). The partial correction scheme outlined in  Appendix~\ref{app:cumulants} only scales the magnitude of the $C_3/C_2$ and $C_4/C_2$
ratios, meaning that the absence of sign alternation and the lack of abrupt variations are expected to persist. 

Note also that even though the temperatures of the freeze-out points in Fig.~\ref{fig:diagram} are relatively high, they will be located below the matching curves when we compare to the HRG models in the next section, so freeze-out curve lies in the HRG regime of the matched model. However, since the matching works smoothly and the freeze-out curve is close to the matching curves, the purely holographic result of Fig.~\ref{fig:qgpcumulants} is a good estimate of the result for the matched model also.

Another way to connect the previously found distinct phase structures to experimental observables is to analyze higher-order baryon number fluctuations near the critical point. We study the ratio $\chifourT/\chitwoT$ of the constrained cumulants, defined using the effective pressure after imposing the $\beta$-equilibrium or HIC constraints in~\eqref{eq:chindef}. This ratio is seen as the proxy of the kurtosis $\kappa$, which according to the above discussion is related to the susceptibilities as $\chi^\mathrm{B}_4/\chi^\mathrm{B}_2 = \kappa \sigma^2$. We study  $\chifourT/\chitwoT$ instead of  $\chi^\mathrm{B}_4/\chi^\mathrm{B}_2$ because it is significantly easier to compute in our setup, but we expect that both observables have qualitatively similar behavior.
The kurtosis has been proposed to be a sensitive probe for locating the CEP in heavy-ion collisions~\cite{Stephanov:2011pb}.

\begin{figure}[!htb]
    \includegraphics[width=0.85\textwidth]{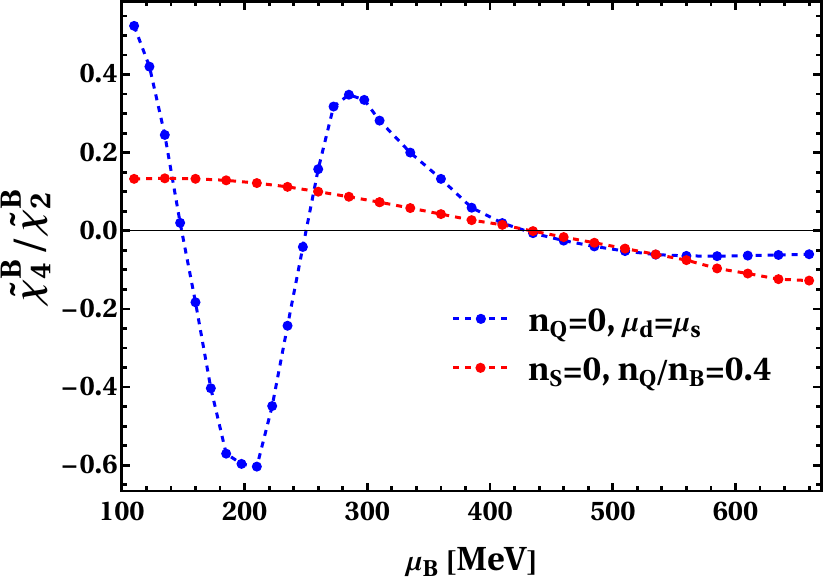}
     \caption{The cumulant ratio $\chifourT/\chitwoT$ for both the $\beta$-equilibrium and HIC setups as a function of the baryon chemical potential. The ratio is evaluated along the trajectories indicated by the blue and red dots in Fig.~\ref{fig:holophasediagram}.}
     \label{fig:C4C2step}
\end{figure}

In Fig.~\ref{fig:C4C2step}, we compare the behavior of $\chifourT/\chitwoT$ across the two setups as a function of the baryon chemical potential along the trajectories shown as the dashed curves in Fig.~\ref{fig:holophasediagram}. In the $\beta$-equilibrated environment, as the thermodynamic trajectory approaches the critical point from the lower-density regime, $\chifourT/\chitwoT$ first dips to negative values before exhibiting a sharp peak. This behavior is qualitatively identical to the standard expectations for critical fluctuations near a CEP. Strikingly, however, at even higher chemical potentials, the ratio drops to negative values again. More importantly, when we evaluate the same observable in the HIC setup, we observe that $\chifourT/\chitwoT$ also becomes negative at large chemical potentials, despite the absence of a critical point or phase transition in this environment. We expect that the kurtosis shows similar behavior in our model, but in order to verify this, we should do the more demanding computation of $\chi^\mathrm{B}_4/\chi^\mathrm{B}_2$ along the same trajectories.

\section{Adding the hadronic phase}\label{sec:heavyion}

In this section, we compare the results of Section~\ref{sec:betaequil} to a hadron gas phase, modeled through hadron resonance gas models. In effect, this will mean replacing the low-temperature parts of the phase diagrams in Fig.~\ref{fig:holophasediagram} by a new phase, which is not coming from a holographic gravity computation but a simple statistical model. Note that the HRG models are known to reproduce the QCD EoS to a high precision within low densities in the low-temperature region. As we shall see, the comparison to HRG models supports the simple interpretation of the phase diagrams in Fig.~\ref{fig:holophasediagram}: in $\beta$-equilibrium the model has a first-order deconfinement transition ending at a critical point, while the transition and the critical point disappear in the HIC setup.

We compute the hadronic thermodynamics using the \texttt{Thermal-FIST} package~\cite{Vovchenko:2019pjl}. Rather than relying on an ideal HRG, we specifically use two interacting variations: the excluded-volume HRG model, EVX \cite{Rischke:1991ke, Vovchenko:2020lju}, and the quantum van der Waals HRG model, vdW \cite{Vovchenko:2015vxa, Vovchenko:2016rkn}. See Appendix~\ref{app:thermalfist} for details. Incorporating repulsive interactions is crucial for dense matter, as excluded volume corrections are known to strongly impact the thermodynamic state and significantly alter the standard chemical freeze-out criteria at high baryon densities~\cite{Cleymans:2006qe}. We selected these models because they match smoothly with the predictions of the flavored V-QCD model at zero baryon density~\cite{Jarvinen:2025mgj}. Therefore using two different models gives a rough estimate on the uncertainty related to the HRG modeling.

We compute the pressure from both HRG models both in $\beta$-equilibrium and for the HIC setup, and compare to the V-QCD results, drawing the combined phase diagrams. We start from the $\beta$-equilibrium case.

\subsection{Charge neutrality and beta equilibrium}\label{sec:HRGvsVQCDbetaeq}

\begin{figure}[!htb]
    \centering
     \includegraphics[width=0.75\textwidth]{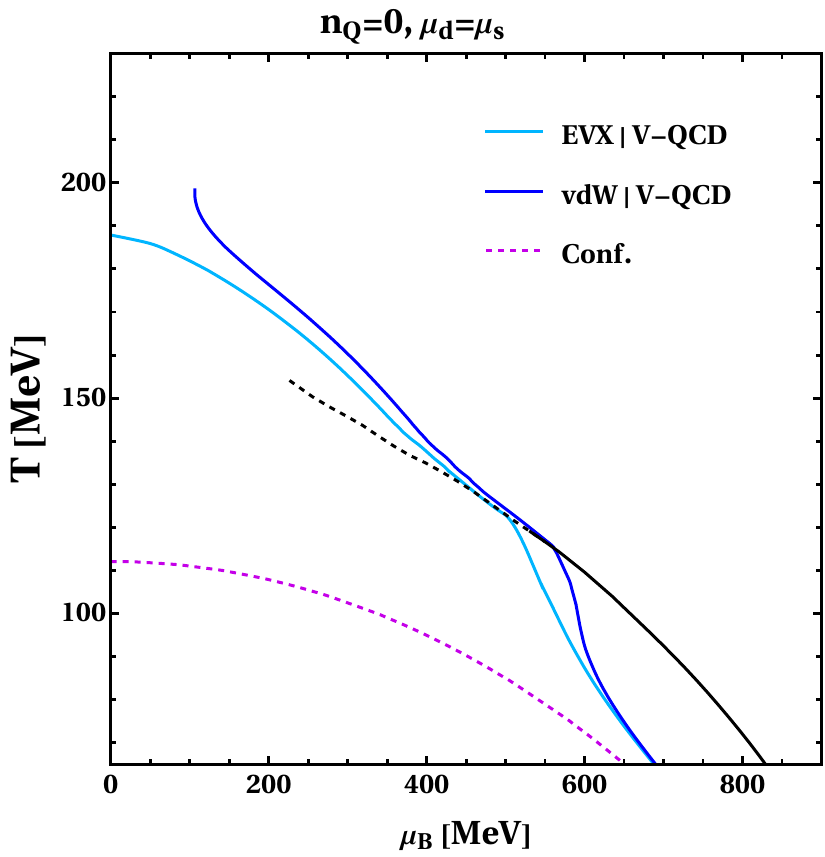}
        \caption{The phase diagram in the ($\muB$,$T$)-plane in $\beta$-equilibrium. 
The black curve shows the first-order transition line between the two black hole phases within the V-QCD model. The light and dark blue correspond to the first order transition lines between the EVX and vdW HRG and the V-QCD model, respectively. The dashed purple curve shows the confinement transition where pressure of the black hole phase becomes negative for this model.}
        \label{fig:pdHRGbeta}
\end{figure}

In Fig.~\ref{fig:pdHRGbeta} the phase diagram of the V-QCD model in $\beta$-equilibrium is shown. The solid lines show the first order transition between the EVX (light blue) and vdW (dark blue) HRG models and the V-QCD result in $\beta$-equilibrium. For the EVX model the coexistence line continues to $\muB=0$. Because vdW model pressure does not intersect with V-QCD at zero density, the vdW coexistence line does not intersect the $\muB$-axis, but continues towards higher temperatures. In principle this line continues to even higher temperatures than shown in the figure and curves towards higher chemical potentials, but we only show the physically reasonable part where the slope of the curve is negative. The black curve shows the transition between black hole phases in V-QCD. The solid part of the curve is the part which is not absorbed by the addition of the HRG, and remains in the final phase diagram, while the dashed part is subdominant to the HRG phases and is removed from the diagram. We also show the confinement transition of the purely holographic solution as the dashed purple line, which is determined by looking where the pressure is zero for the black hole solution. However, note that this transition happens in between phases that are subdominant after the HRG models are added, so it is not a part of the final phase diagram. That is, only the black solid curve and either of the blue solid curves (depending on the choice of the HRG model) are present in the final diagram.

\begin{figure}[!htb]
    \includegraphics[width=0.5\textwidth]{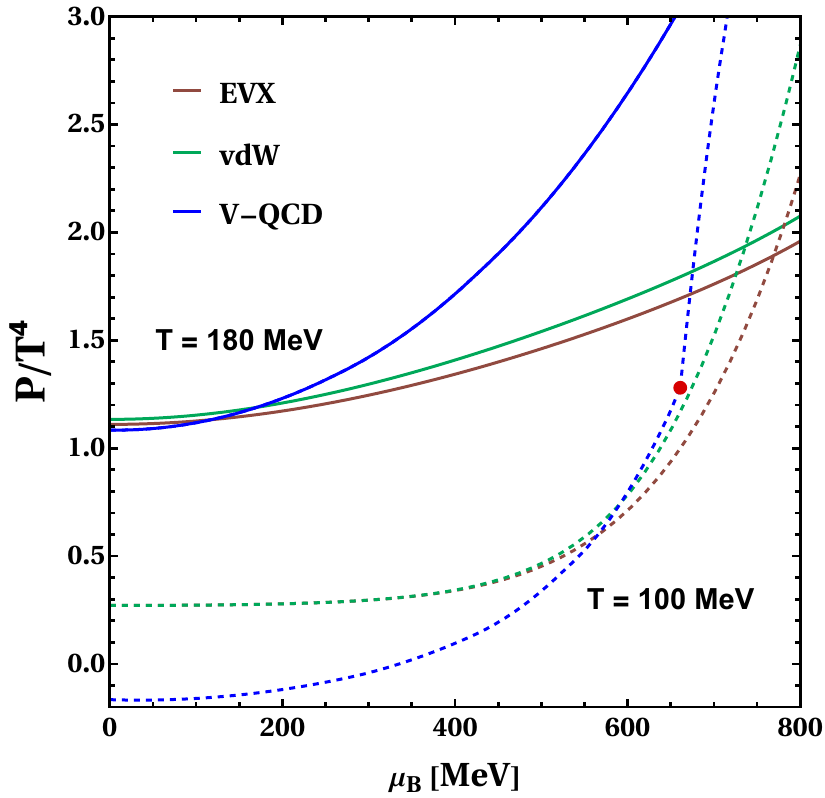}
    \includegraphics[width=0.5\textwidth]{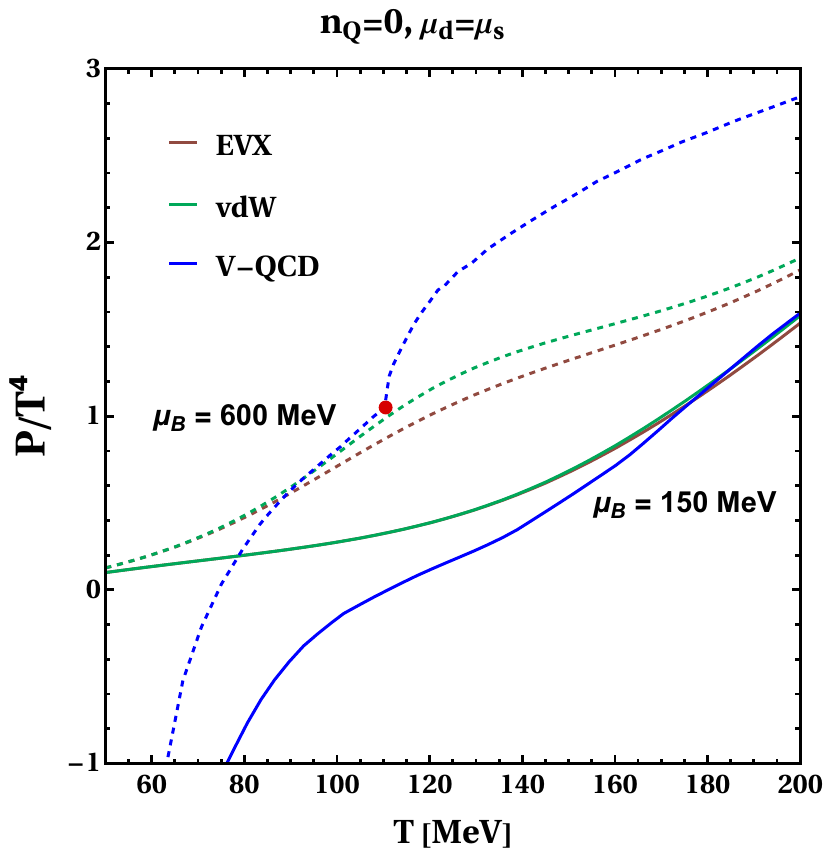}
     \caption{The normalized pressure $P/T^4$ as a function of the chemical potential $\muB$ (left) at fixed temperature and as a function of the temperature $T$ at fixed chemical potential (right) for the EVX HRG (red), vdW HRG (green), and $\beta$-equilibrated V-QCD (blue) models. The purely holographic V-QCD phase transition is shown as the red dot. }
     \label{fig:pressurebeta}
\end{figure}

In order to provide more details about the equation of state, we show the normalized pressure $P/T^4$ of V-QCD (blue curves) in $\beta$-equilibrium compared to the HRG models (green and red curves) in Fig.~\ref{fig:pressurebeta}.
The left panel shows $P/T^4$ as a function of $\muB$ at two different values of the temperature, while the right panel shows the same quantity as a function of the temperature at two values of the chemical potential.
The purely holographic phase transition between two kinds of black hole solutions in the V-QCD model is shown as the red dot. The values of temperature and chemical potential in each panel were chosen so that one has the phase transition and one does not. In these plots we only show the curves corresponding to the dominant phase in the purely holographic V-QCD picture, and around the phase transition points there are also subdominant phases. The final pressure for each choice of the HRG model is obtained by following the curves with highest pressure.

For the V-QCD curves with phase transition, the parts of the curves corresponding to the low-temperature black hole phase (sections left of the red dot in both panels) seem to match quite well with the HRG models, in particular the vdW model, near the phase transition. However, the sections corresponding to the high temperature black hole phase (sections right of the red dot in both panels) do not match with the HRG results.

\begin{figure}[!htb]
    \includegraphics[width=0.433\textwidth]{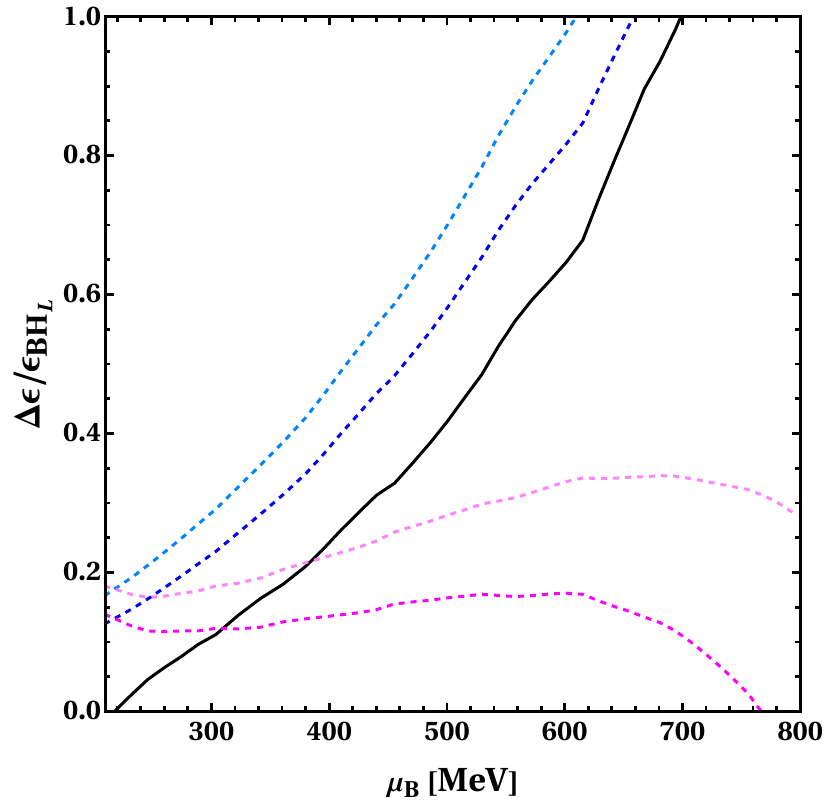}
    \includegraphics[width=0.55\textwidth]{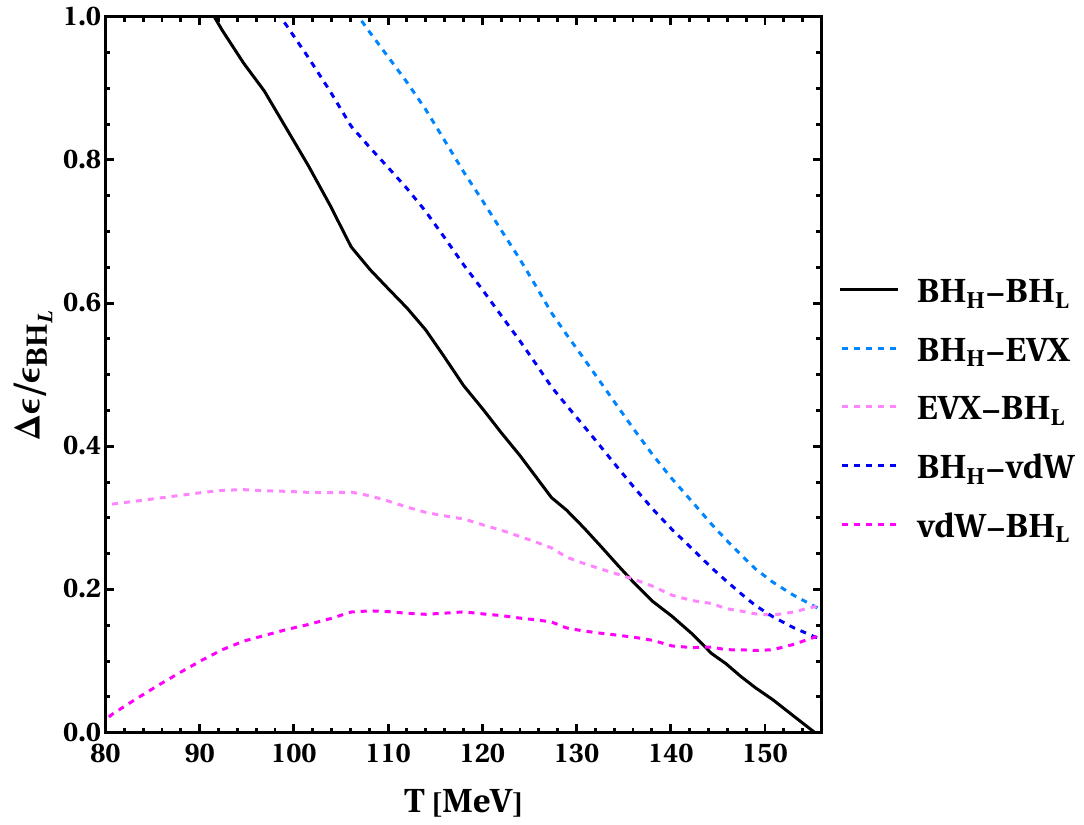}
    \caption{The change in energy density at the black hole phase transition of $\beta$-equilibrated V-QCD  as a function of the chemical potential $\muB$ (left) and temperature $T$ (right) normalized by the energy density of the $\text{BH}_L$ phase. The black curve is the latent heat across the two black hole phases and it goes to zero as we approach the critical point. The various colors show the energy density difference at the black hole phase transition between the HRG models and the two black hole phases in V-QCD.}
    \label{fig:Ltrans}
\end{figure}

This phenomenon may be studied more closely by plotting  the change in energy density between various phases along the purely holographic phase transition, \ie, the black curve in Fig.~\ref{fig:pdHRGbeta}. We show the results in Fig.~\ref{fig:Ltrans}.
The black curve is the actual (normalized) latent heat, \ie, the relative difference in energy density between the two different V-QCD black hole phases, which goes to zero as we move towards the critical point. The dashed lines show the energy differences across other phases present at the same transition line. All curves are normalized to the energy density of the low-temperature V-QCD black hole phase. 

The pink curves show the relative energy difference between the low-temperature V-QCD phase and the HRG models, with the EVX (vdW) model shown as the light (dark) curve. We observe that this difference is relatively low over the whole phase transition curve.  The energy density difference between the high-temperature V-QCD phase and the HRG phases, shown as the blue curves, with the EVX (vdW) HRG model corresponding to the light (dark) curve. In stark contrast with the curves of the low temperature V-QCD phase, these differences grow strongly with increasing chemical potential or decreasing temperature. This shows that, in agreement with our observations from Fig.~\ref{fig:pressurebeta}, the HRG phases match well with the low-temperature V-QCD black hole phase near the critical line, while there is no agreement between the high-temperature V-QCD phase and the HRG models. The agreement between the V-QCD low-temperature phase and the vdW model is particularly good for the whole range of chemical potentials. Our interpretation of this is that the low-$T$ V-QCD and HRG phases in the matched model should be understood as a model of the same physical phase. That is, even though the low-temperature V-QCD black hole phase is in principle deconfined by holographic dictionary, it acts as a proxy of the confined hadron gas phase. Therefore, the purely holographic transition observed in the left panel of Fig.~\ref{fig:holophasediagram} in Section~\ref{sec:betaequil} can be interpreted as the deconfinement transition. However, recall that according to the analysis in Section~\ref{sec:betaequil} (see Fig.~\ref{fig:tau}) the horizon value of the strange quark tachyon components drops significantly as one crosses the transition from low to high temperatures. Except in the immediate vicinity of the CEP, the high-temperature value is much less than one. Therefore the chiral symmetry in the strange quark sector is also restored at the transition, up to a suppressed explicit breaking effect due to the finite quark mass, so that the transition should actually be interpreted as a combined deconfinement and chiral restoration transition.

\begin{figure}[!htb]
    \includegraphics[width=0.5\textwidth]{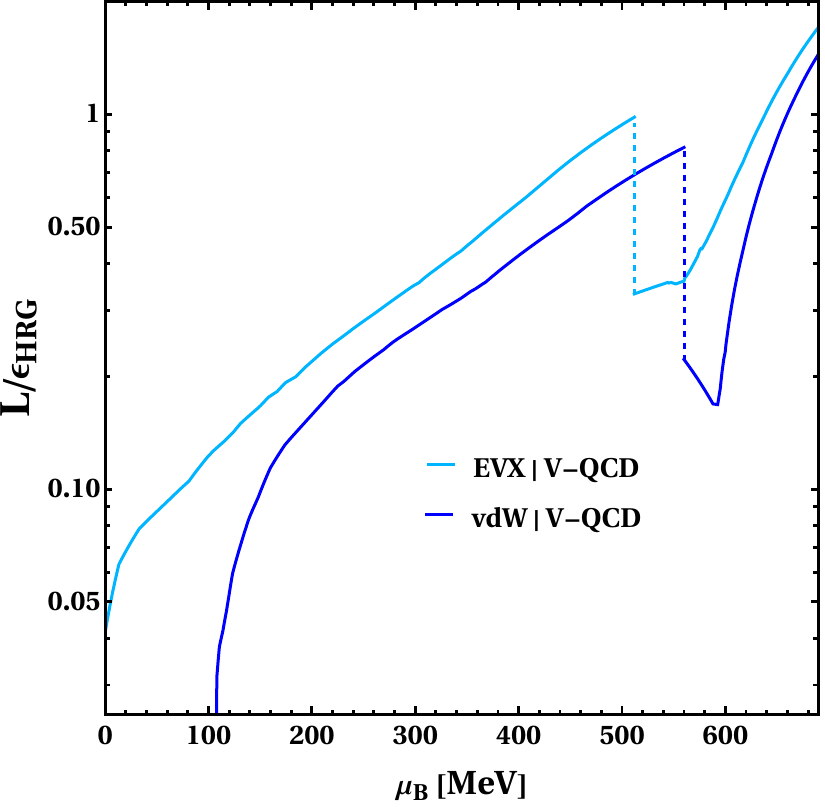}
    \includegraphics[width=0.5\textwidth]{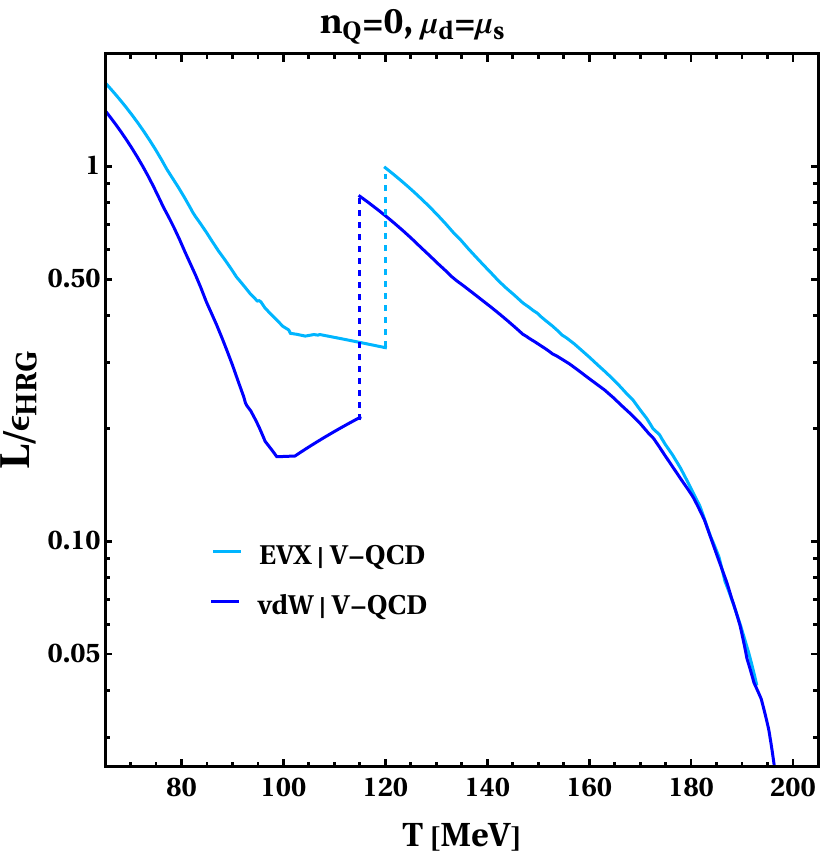}
    \caption{The latent heat at the phase transitions between the HRG and V-QCD models in $\beta$-equilibrium as a function of the chemical potential $\muB$ (left) and temperature $T$ (right). The vertical axis is logarithmic and normalized by the HRG energy density. In the temperature plot the line for the EVX model is truncated as the phase transition curve intersects the $\muB=0$ axis at this temperature. The light blue and dark blue dashed lines show the discontinuity in the latent heat across the black hole phase transition.}
    \label{fig:Lbeta}
\end{figure}

In order to obtain further confirmation for our interpretation, we show the latent heat across the HRG to V-QCD phase transitions (shown as blue curves in Fig.~\ref{fig:pdHRGbeta}) in Fig.~\ref{fig:Lbeta}.
The left (right) panel shows the normalized latent heat as a function of the chemical potential (temperature) along the transition lines. Looking at the left panel, the phase transition starts off weak at small chemical potentials. As the chemical potential grows, $\muB \gtrsim 300$~MeV, the normalized latent heat grows and becomes $\mathcal{O}(1)$. However, at around $\muB\approx 500$~MeV, the latent heat drops and the transition becomes weak again. This drop marks the spot where the blue HRG to V-QCD transition curves in the phase diagram of Fig.~\ref{fig:pdHRGbeta} cross the black curve of the purely holographic phase transition. That is, in full agreement with our observations from Figs.~\ref{fig:pressurebeta} and~\ref{fig:Ltrans}, when the phase transition becomes that between the low-temperature V-QCD phase and the HRG phase instead of the high-temperature V-QCD phase, the phase transition becomes weak. At even higher chemical potentials above $\muB \approx 600$~MeV, the latent heat grows and the phase transition becomes strong again. However, as seen from Fig.~\ref{fig:pdHRGbeta}, at these values of the chemical potential the transition enters the region of such low temperatures and high $\muB$ that contributions from nuclear matter may become significant. Since our modeling does not contain nuclear matter, the results in this regime may not be trustworthy.

In conclusion, strong first-order phase transitions in Fig.~\ref{fig:pdHRGbeta} are found on the black solid curve and on the sections of the blue curves above the black dashed curves, except in the low chemical potential region $\muB \lesssim 300$~MeV, where the phase transitions shown as the blue curves become weak. Therefore, one can interpret the combined strong first-order transition curve arising from these sections of the solid black and blue curves in the combined phase diagram of V-QCD and HRG as the proxy of the FOPT of the purely holographic computation (combination of the black solid and dashed curves in this figure).  However, the matching of the HRG and V-QCD low-temperature phases is not precise enough to pinpoint accurately the location of the critical point in the matched model. Nevertheless, the results of the matched model can be interpreted as consistent with the critical point in the same region as where it is found in the purely holographic computation of Fig.~\ref{fig:holophasediagram}. Comparison between the dashed black curve and the solid blue curves in Fig.~\ref{fig:pdHRGbeta} suggests that the critical temperature moves to slightly higher values in the matched model compared to the purely holographic analysis in the near-critical region. However, because the matching between the HRG and V-QCD models is rather delicate as seen from the pressure comparison in Fig.~\ref{fig:pressurebeta}, the precise location of the curves may be sensitive to the way we compare the V-QCD model to lattice data (see Appendix~\ref{app:cumulants}) or even variations within the error bars of the lattice data.

Finally, as we stated above, we are interpreting the FOPT of the final matched model as a combined deconfinement and chiral phase transition. Recall that in the purely holographic computation the confinement transition was the transition between horizonless and black hole phases, marked as the dashed purple curve in Fig.~\ref{fig:pdHRGbeta}, whereas the black curve was the transition between black holes and would only involve (partial) restoration of the chiral symmetry. The interpretation in the matched phase diagram is that the addition of the hadron gas pressure component through the HRG models pushes the confinement transition to higher temperatures, close to the line of the black-hole transition in the purely holographic model, therefore leading to a combined deconfinement and chiral transition. This is similar to what was seen in earlier comparisons of the unflavored V-QCD models with the HRG pressure in~\cite{Demircik:2021zll,Ecker:2025vnb}, except for the behavior of the critical point: here we see that the critical point emerges already in the purely holographic analysis, whereas in the unflavored analysis the critical point of the physical phase diagram\footnote{Recall that as we pointed out in Section~\ref{sec:envi}, a  transition ending in a critical point can also be found in the purely holographic unflavored V-QCD setup but it appears only between black hole phases that are thermodynamically subdominant to the confined phase.} could only be identified after comparing the results from the V-QCD model with hadron gas pressure.

\subsection{Heavy-ion conditions}\label{sec:HRGvsVQCDHIC}

We then analyze the matching of the V-QCD model with the HRG models in the other physical framework considered in this article, \ie, the heavy ion conditions keeping $n_\mathrm{S} = 0$ and $\nQ/\nB=0.4$.

\begin{figure}[!htb]
    \centering
     \includegraphics[width=1.1\textwidth]{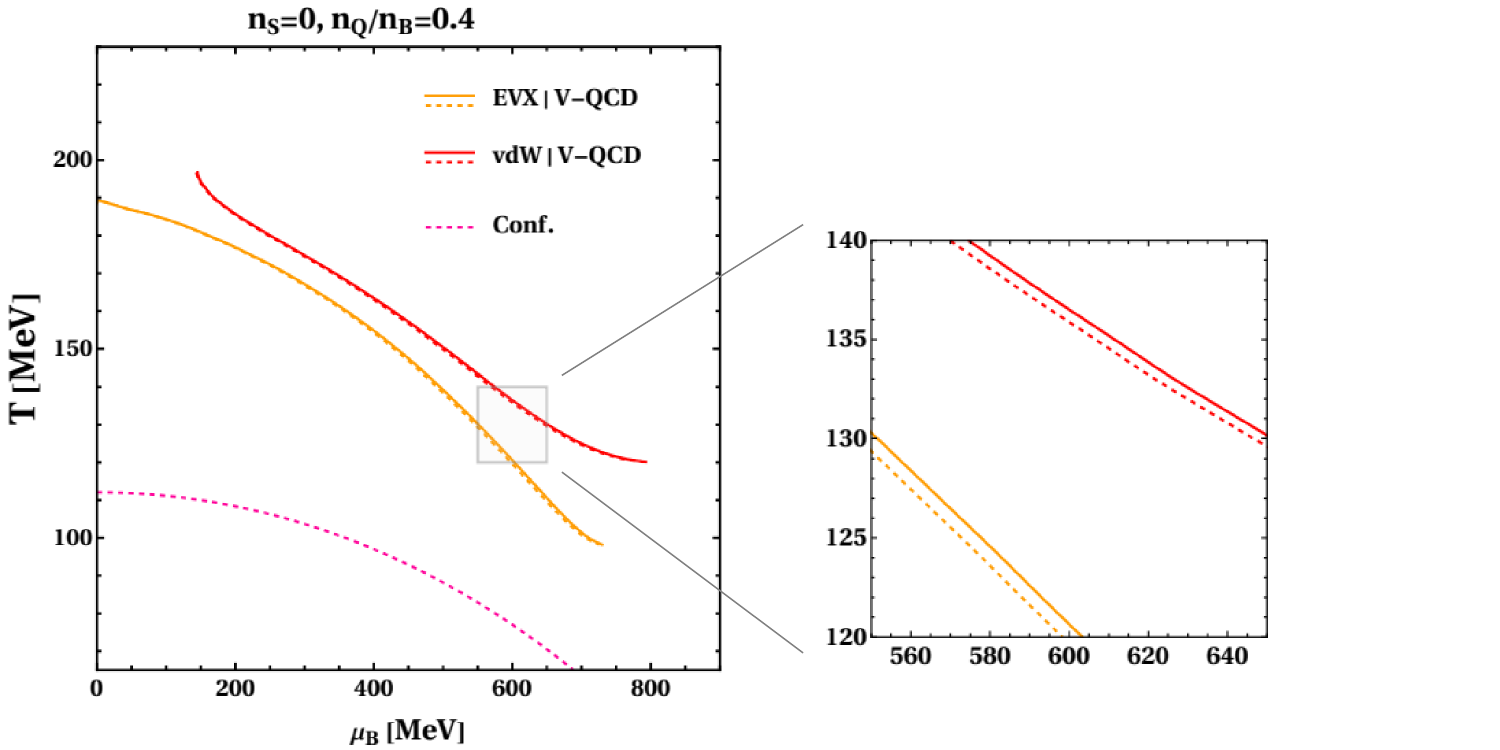}
        \caption{The phase diagram in the ($\muB$,$T$)-plane for V-QCD model at heavy-ion conditions (zero strangeness and $\nQ = 0.4 \,\nB$). The yellow (red) curves correspond to the first-order transition lines between the EVX (vdW) HRG and the V-QCD model, respectively. Due to the choice of ensemble (see text), the phase transition splits into two curves on the $(\muB,T)$-plane, which are shown as the dashed and solid curves. The splitting is better visible on the right plot where we zoom in to the phase transition region. }
        \label{fig:pdHRGQGP}
\end{figure}

In Fig.~\ref{fig:pdHRGQGP}, the phase diagram of the combined model after matching the V-QCD and HRG pressures is shown. Recall that the purely holographic phase diagram in Fig.~\ref{fig:holophasediagram} (right) only has one transition, the deconfinement transition marked in Fig.~\ref{fig:pdHRGQGP} as the dashed purple curve. This transition is removed after matching as the HRG models are dominant at low temperature and density. The only transition in the final phase diagrams are solid and dashed yellow and red lines, which show the first-order transition between the EVX and vdW HRG models and the V-QCD model. Recall that the coexistence line is determined by matching the chemical potential $\tilde \mu = \muB + 0.4 \,\muQ$ in~\eqref{eq:mueff}. Therefore $\muB$ is not the same between the phases at equilibrium, so on the $(\muB,T)$-plane we obtain two lines. However, note that the difference between the dashed and solid lines is highly suppressed. The region between the solid and dashed lines is in principle filled by the coexistence or mixed phase of the ensemble defined in terms of $\tilde \mu$. For the EVX model the coexistence line continues all the way to $\muB=0$. As was the case in $\beta$-equilibrium in Fig.~\ref{fig:pdHRGbeta}, for the vdW model the coexistence line does not intersect the $\muB=0$ axis, but continues towards higher temperatures. We again only show the transition in the range where the slope of the transition line is negative. In both models at large chemical potential the phase transition starts to curve towards higher temperatures and hence the lines are truncated in both models. However, this happens at such high chemical potentials that both the HRG models (being a low temperature low density description) and the V-QCD model (which here does not include nuclear matter) may become unreliable. 

\begin{figure}[!htb]
    \includegraphics[width=0.5\textwidth]{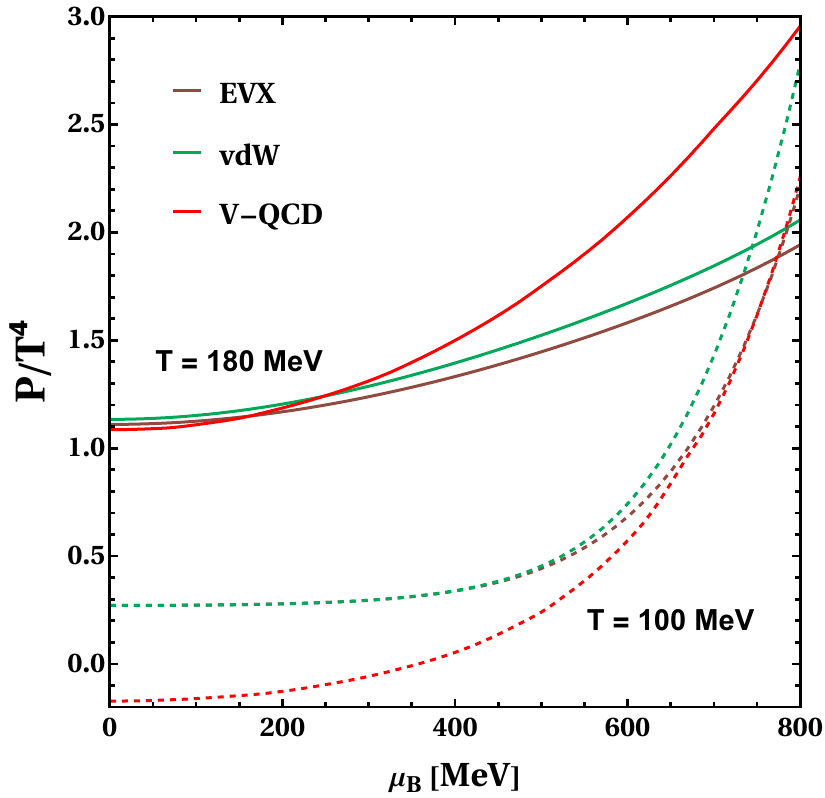}
    \includegraphics[width=0.5\textwidth]{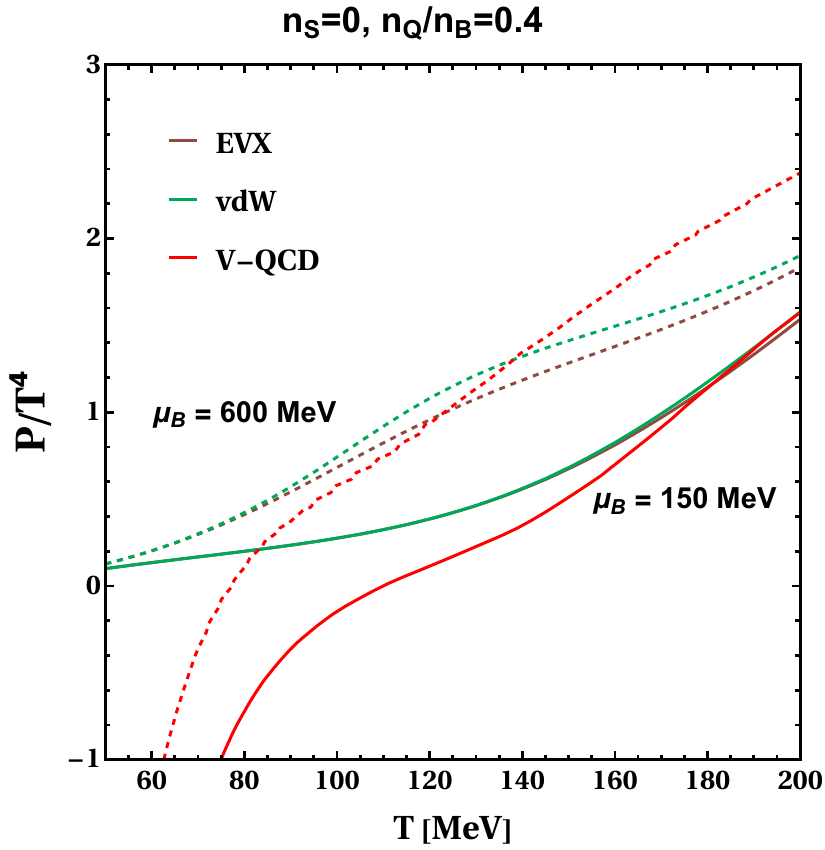}
    \caption{The pressure as a function of the chemical potential $\muB$ (left) and temperature $T$ (right) for the EVX, vdW and V-QCD model. The pressure is shown for two parameter values, which are $T=100 \ \text{MeV}$ and $T = 180 \ \text{MeV}$ for the chemical potential plot and $\muB=600 \ \text{MeV}$ and  $\muB=150 \ \text{MeV}$ for the temperature plot. These are distinguished by dashed and solid lines and the corresponding curves are indicated in the figure.}
    \label{fig:pressureqgp}
\end{figure}

In order to give more details, we show the normalized pressure $P/T^4$ of the V-QCD model in comparison to the pressure of the HRG models in Fig.~\ref{fig:pressureqgp}. The conventions in these figures follow Fig.~\ref{fig:pressurebeta} which showed similar comparison in $\beta$-equilibrium. The final result after matching is obtained by taking the highest (normalized) pressure at each parameter value, after comparing between the V-QCD model and either of the HRG models. We note that in each of the constant temperature or chemical potential slices presented in this figure, the pressure of the V-QCD model matches relatively smoothly with the HRG curves near the location of the phase transition where the pressure curves intersect.\footnote{Except for the vdW model at $T=100$~MeV, where no intersection is present.}

\begin{figure}[!htb]
    \includegraphics[width=0.5\textwidth]{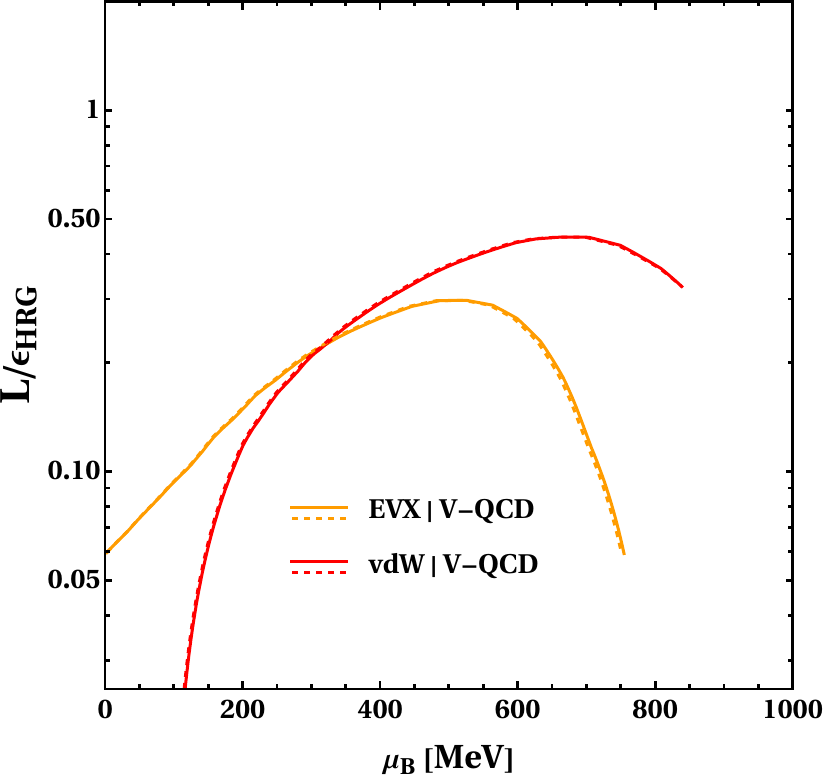}
    \includegraphics[width=0.485\textwidth]{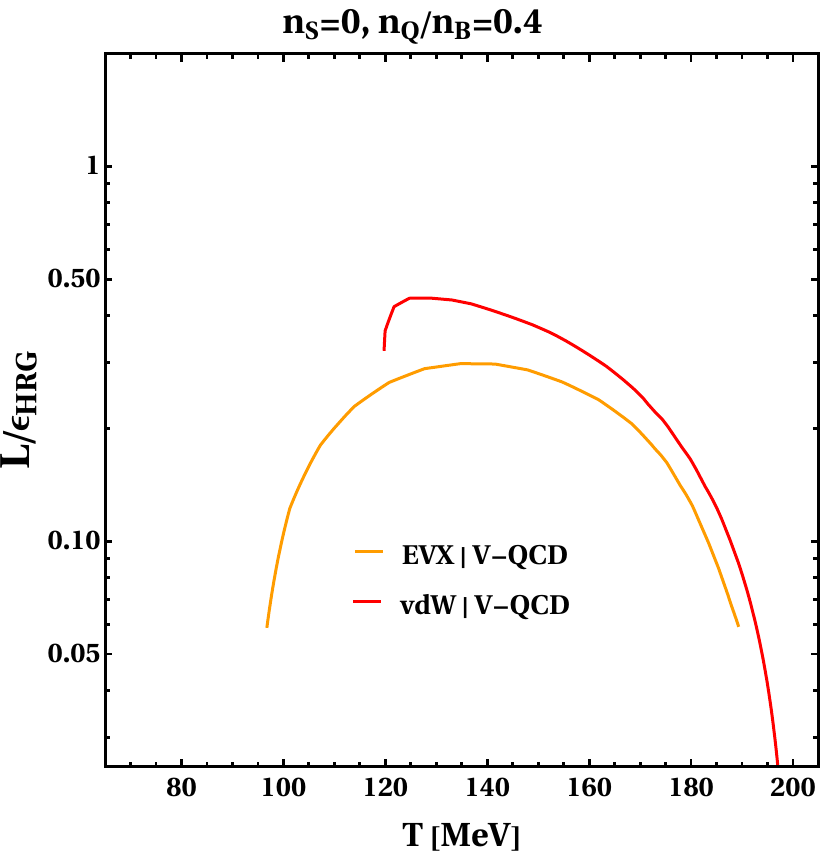}
    \caption{The latent heat at the HRG phase transitions for the V-QCD model as a function of the chemical potential $\muB$ (left) and temperature $T$ (right). The vertical axis is logarithmic and normalized by the HRG energy density. The latent heat is single-valued at each transition point, but it gives rise to two different curves when projected onto the baryon chemical potential $\muB$ of either phase. Here dashed corresponds to HRG and solid to V-QCD. In the temperature plot there is only one line. The truncation of the lines corresponds to how the phase transitions behave in the phase diagram Fig.~\ref{fig:pdHRGQGP}.}
    \label{fig:Lqgp}
\end{figure}

In order to study the matching between the V-QCD and the HRG models quantitatively, we show the latent heat across the transitions of Fig.~\ref{fig:pdHRGQGP} as a function of baryon chemical potential and temperature in Fig.~\ref{fig:Lqgp}. The solid and dashed lines correspond to the same curves as shown in the phase diagram. The latent heat is normalized by the energy density of the HRG model in question. Note that the latent heats are only shown in the same ranges as where we show the transitions in the phase diagram of Fig.~\ref{fig:pdHRGQGP}.  We see in both cases the numerical values are relatively low so the transition is weak, similarly to the differences between the energy density and latent heat between the HRG models and the low-temperature V-QCD black hole phase in the $\beta$-equilibrium scenario in Figs.~\ref{fig:Ltrans} and~\ref{fig:Lbeta}.

Our interpretation of this result is that the weak transition should be understood as the proxy of a crossover from the confined phase to the deconfined phase. Therefore there is no critical point or a physical phase transition in the phase diagram for the heavy-ion conditions, at least in the range of parameters where our construction is expected to be reliable, confirming the expectation from the purely holographic analysis of Fig.~\ref{fig:holophasediagram} (right). This interpretation is similar to that obtained when matching the unflavored V-QCD model with the HRG model in ~\cite{Demircik:2021zll,Ecker:2025vnb} where a weak FOPT was also interpreted as a crossover at low chemical potentials. A more detailed analysis with added contributions from nuclear matter might reveal a CEP in the high-density regime, well above $\muB \approx 600$~MeV, in the same way as in the unflavored case of~\cite{Demircik:2021zll,Ecker:2025vnb}. However, it is not obvious that pushing the heavy-ion model to such high chemical potentials makes sense.

\section{Discussion}\label{sec:discussion}

In this work, we discussed the phase diagram of hot and dense QCD matter by using a flavor-dependent holographic model. To this end, we used the bottom-up holographic V-QCD framework,  matched directly to flavor-dependent lattice thermodynamics at zero density, to analyze the equation of state of the deconfined QCD phase. This implementation explicitly incorporates a realistic quark mass hierarchy where up and down quarks are kept massless and the strange quark is massive. We started by analyzing the predictions directly from the holographic model, \ie, in the pure dual gravity picture, in Section~\ref{sec:betaequil}. We went on matching the results from the high-temperature phase of the holographic model with low-temperature results from hadron resonance gas models in Section~\ref{sec:heavyion}. 

In the purely holographic picture, applying this single, unified V-QCD model to different physical environments reveals a sharp divergence in the resulting dense-matter phase structure. Under the charge-neutral, $\beta$-equilibrated conditions typical of neutron stars, the model predicts a first-order phase transition that terminates at a critical endpoint. In contrast, this phase transition disappears entirely under the conditions of strangeness neutrality $\nS=0$ and a fixed charge-to-baryon ratio $\nQ/\nB=0.4$ dictated by heavy-ion collisions.   

We then checked whether a clear picture emerges if the low-temperature part of the phase diagram is replaced by the hadron resonance gas, which we expect to give a more fundamental description of the EoS in this region than the holographic model. This was in part motivated by earlier studies~\cite{Demircik:2021zll}, where the gravity computation did not show a critical point, but after matching with the hadron resonance gas result, a CEP would emerge. By analyzing the phase diagram and jumps in the energy density at matching, we noticed that smooth matching between the HRG and V-QCD results was possible for a wide range of chemical potentials,  from zero up to $\muB \approx 600$~MeV ($800$~MeV) in the $\beta$-equilibrium (HIC) setup, respectively. This is remarkable given the relative simplicity of the HRG setup and the crudeness of the fit to lattice data for susceptibilities in V-QCD. To be precise, when using the $\beta$-equilibrium conditions, the HRG results match well with the low-temperature gravity phase (BH${}_\mathrm{L}$) near the phase transition line of the pure gravity model, suggesting that this low-temperature phase can be identified with the confined hadron gas phase,  whereas the high-temperature black hole phase is still separated by a clear first-order transition. Analyzing the latent heats of the transition curves suggests that a CEP is found in the matched model in the same region as where it is found in the pure gravity computation. While the precise location of the critical point is unclear in the model matched with HRG, the transition at higher densities is clear, identified as a confinement-deconfinement transition, and the phase transition curve is essentially the same curve as obtained in pure gravity. In the HIC setup the picture is simpler: the HRG EoS matches relatively smoothly with the single deconfined V-QCD phase in a large range of densities, and therefore no sign of critical points is found even after matching.

All the phase transitions considered in this paper are of the congruent type, meaning that the macroscopic density conditions are imposed on both the deconfined and hadronic phases independently. This includes the black hole transition found when the V-QCD model is subjected to $\beta$-equilibrated conditions. Because a congruent construction forces the two phases to maintain identical chemical compositions, it is generally not possible to match all the individual chemical potentials at the phase interface. To achieve a complete matching of all chemical potentials, one would instead need to use a non-congruent phase transition framework~\cite{Iosilevskiy:2010qr,Tatsumi:2011tt,Hempel:2013tfa}. This allows for the formation of a mixed phase where the two competing states possess distinct local charge densities that only satisfy the physical conditions on average. Choosing a congruent setup over a non-congruent one structurally impacts the global features of the phase diagram and the exact layout of the coexistence regions.

In this work we restricted our analysis to two constrained scenarios, given by the $\beta$-equilibrium and heavy-ion conditions, which reduce the phase diagram to a two-dimensional space conveniently parameterized by the temperature and the baryon number chemical potential. It would be interesting to carry out a more exhaustive study of the phase diagram, which in general depends on three chemical potentials. Such a more exhaustive study could reveal, among other things, how the critical point disappears as one moves from the $\beta$-equilibrium conditions towards the heavy-ion conditions.

Our study ignored one important point of the gravity modeling of the phase transition and the critical point, namely, the inhomogeneous instability analyzed in~\cite{CruzRojas:2024igr,Demircik:2024aig}. While the instability was shown in~\cite{Demircik:2024aig} to be unavoidable in flavor-independent models that are fitted to baryon number susceptibilities, the situation in the flavor-dependent framework is currently unclear. The question is complicated by technical issues. First, turning on the strange quark mass implies that chiral symmetry is broken in all phases. The instability is driven by Chern--Simons terms in holographic QCD, and while such terms were solved very recently in the chirally broken phase~\cite{Raymond:2026twp}, their structure is complicated. Second, turning on the strange quark mass means that the holographic background is non-Abelian, \ie, the background solutions for the tachyon and the gauge fields can no longer be taken to be proportional to the unit matrix. In this case, the V-QCD fluctuation equations, which one needs to analyze to locate the instability, become ambiguous due to the use of the DBI action. However, the ambiguity does not affect the EMD setup. This also suggests that at low densities, where the DBI and EMD descriptions become equivalent~\cite{Jokela:2024xgz}, the ambiguity is suppressed, and the instability can be mapped reliably also in V-QCD.

Our findings also suggest other directions for systematically improving the bottom-up holographic description of dense QCD matter. First, the stark dependence of the phase structure on the specific physical environment demonstrates the severe limitations of frameworks that are fitted solely to bulk baryon number susceptibilities. To move past this limitation, such frameworks must be generalized to include realistic flavor physics. Following the approach of our work, this can be achieved by introducing gauge fields transforming under the full flavor symmetry and flavored complex scalar fields dual to quark bilinears, thereby capturing explicit quark mass effects and flavor dependencies without requiring a full top-down string theoretic completion.

In this work we used the V-QCD model, but we expect that essentially the same or similar results as in V-QCD (\eg, for the location of the CEP) will be found in EMD setups generalized to include flavors, \ie, Einstein--Yang--Mills-dilaton models with additional complex scalar fields fitted to lattice results for  quark susceptibility matrix. Similarity of the results is expected because, as seen in~\cite{Jokela:2024xgz}, changing the flavor action from DBI to Yang--Mills affects the results surprisingly little when chiral symmetry is intact. However, it is a priori not clear that the same holds after turning on the strange quark mass, even if its value is rather small. We do expect that tachyon dependence analogous to the dependence of the $w$ function in~\eqref{eq:wdef} is required also when Yang--Mills action is used for flavors. Moreover, one can see from Fig.~\ref{fig:tau} that the value of the tachyon at the horizon (which is usually the maximal value of the tachyon in the solution) satisfies $\tau_h < 0.2$ near the critical point. As the action only depends on even powers of the tachyon field, this signals that nonlinearities in the tachyon are small. In the absence of nonlinearities, the V-QCD and EMD setups (after the addition of a complex scalar field controlling the chiral symmetry and playing the role of the tachyon) should be essentially equivalent.\footnote{We have also carried out some preliminary checks in the V-QCD setup, changing the definition~\eqref{eq:wdef} at nonlinear level in the tachyon. These checks suggest that our results are insensitive to such modifications, providing further support to the expectation that similar results are found for the Yang--Mills action.} Despite this, it was observed in~\cite{Jarvinen:2025mgj} that nonlinearities of the tachyon do play a minor role in the data fitting procedure, which we also use in this article. Therefore, while we expect that the strange quark mass effects in the EMD setups are similar to the V-QCD analysis, this should be verified by explicit computation.

The flavored fitting procedure presented here had a major limitation: the flavor-diagonal nature of our underlying action forces the off-diagonal quark susceptibilities, $\chi_{ij}$ with $i\ne j$, to vanish at zero density. Consequently, we were able to do a good fit to lattice data for quark susceptibilities, but failed to fit the baryon number susceptibilities. This leads to the systematic underestimation of the higher-order net-proton cumulant $C_3/C_2$ along the chemical freeze-out curve when compared against experimental data from the RHIC BES. To properly account for these cross-quark correlations, future studies should explore promoting the flavor action to its fully non-Abelian version, a demanding task that requires tracking ``non-coincident'' D-branes~\cite{Erdmenger:2007vj,Herzog:2008bp,Hoyos:2025qwd}, or implementing multitrace boundary operators that avoid $1/N$ suppression in the Veneziano limit. Perhaps the most natural framework for doing this would be Einstein-Yang--Mills-dilaton models, because defining the multitrace generalization of the DBI action in the V-QCD model is not straightforward.

Complementary information regarding these off-diagonal susceptibilities can also be obtained from perturbative QCD (pQCD). At both vanishing and finite baryon chemical potential, the leading contribution to flavor mixing arises at order $g^6\log(1/g)$ in the strong gauge coupling $g$. This represents an $\mathcal{O}(\alpha_s^3)$ or next-to-next-to-next-to-leading-order calculation in terms of the strong coupling constant $\alpha_s\equiv g^2/(4\pi)$. This contribution is driven entirely by a single four-loop process: `the Bugblatter' diagram, which features two distinct quark loops connected by three gluon exchanges. At $\muB=0$, these flavor-mixing susceptibilities were first computed in~\cite{Blaizot:2001vr,Blaizot:2002xz}, with the complementary diagonal components subsequently evaluated at the same order~\cite{Vuorinen:2002ue}. Recent advancements in (dense) loop-tree duality methods~\cite{Navarrete:2024zgz} now allow for the direct numerical evaluation of this complex topology at finite density. At $\muB=0$, the UV behavior of the V-QCD action is already constrained by qualitative QCD physics and matched to explicit two-loop perturbative results. Extending this approach to match the model against these state-of-the-art finite-density pQCD calculations is in the same spirit. Such a quantitative comparison would provide valuable insight into how the holographic flavor action can be refined in the deep UV to better capture the underlying physics of dense quark matter.

Recall also that in this work we neglected the V-QCD black hole phase where chiral symmetry is fully broken and also the light quark tachyon components condense in the bulk. Such a phase was seen to exist at zero density in~\cite{Jarvinen:2025mgj} but it was limited to low temperatures as compared to the region of focus in this article. Moreover, we argued that the low-temperature V-QCD result cannot be used as such in a realistic EoS model for QCD, as it misses meson loop contributions. Therefore, considering the details of the pure V-QCD phase diagram at low-temperature might not sound interesting. Nevertheless it would be worthwhile to check what happens to this phase at nonzero density: Comparison with HRG suggests that, while in purely holographic V-QCD computation chiral symmetry in the strange quark sector is restored (neglecting the small effect due to the strange quark mass) at the transition between the low temperature and high temperature black holes, in the matched model also the chiral symmetry of the light quarks is restored at the same transition. It would be interesting to see if a similar symmetry restoration pattern can be observed in the purely holographic computation after adding the chirally broken black hole phase, \ie, if the light quark chiral transition also takes place along the same transition curve at least at high densities. This would mean improved agreement between the pure V-QCD model and the model obtained after matching with HRG. However, the numerical analysis of the chirally broken black hole solutions is challenging: it would require either drastically optimizing the numerical method used in this article, solving the geometries using a different method which is more efficient for the chirally broken backgrounds, or significantly more computational resources than the analysis carried out in this article.

Finally, there is also ongoing work aiming to extend the flavored model towards the regime of even higher densities and low temperatures, in order to analyze the properties of neutron stars~\cite{Jarvinen:2021jbd,Hoyos:2021uff}. This requires implementing nuclear matter in the model, \eg, following ideas from~\cite{Ishii:2019gta,Demircik:2021zll,Bartolini:2025sag,Ecker:2025sjb} and potentially also other (such as color superconducting~\cite{BitaghsirFadafan:2018iqr,Henriksson:2019zph,BitaghsirFadafan:2020otb,Henriksson:2022mgq,Preau:2025ubr,CruzRojas:2025fzs})  phases.

\section*{Acknowledgments}
We thank Tuna Demircik, Christian Ecker, Wanxiang Fan, Song He, Carlos Hoyos, Mei Huang, Romuald Janik, Elias Kiritsis, Li Li, Andrea Olzi, Risto Paatelainen, Edwan Pr\'eau, Yi-Ping Si, Javier Subils, Aleksi Vuorinen, and Hong-An Zeng  for discussions. N.~J. was supported in part by the Research Council of Finland through grant no. 354533 and the Centre of Excellence in Neutron-Star Physics (project 374062). T.~M. has been supported by the DFG through the Emmy Noether Programme (project
number 496831614), through CRC 1225 ISOQUANT (project number 27381115).

\appendix

\section{Potential sets} \label{app:st_potsform}
In this Appendix we present the expressions for the potentials we use in this article. We present the Ansatz and the parameters given by the comparison to the lattice data.

As explained in the main text, we set $\Nc=3$ and $\Nf=3$.
The complete Ansatz for the potentials is given by 
\begin{align}
 V_\mt{g}(\lambda) & =12\,\biggl[1+V_{g,1} \l+{V_{g,2}\lambda^2
\over 1+c_\l \l/\l_0}+V_\mathrm{IR} e^{-\l_0/(c_\l\l)}\left(\frac{c_\l\l}{\l_0}\right)^{4/3}\!\!\sqrt{\log(1+c_\l\lambda/\l_0)}\biggr]\  \\
\label{Vfan}
 V_\mt{f}(\lambda,\tau_i) & = V_{\mt{f}\lambda}(\l) \left(1+
\tau_i^4\right)^{\tau_p} \exp\left(-\tau_i^2\right)   \\
V_{\mt{f}\lambda}(\l) & = W_0 + W_1 \l +\frac{W_2 \l^2}{1+c_\lambda\l/\l_0} + 12 W_\mathrm{IR}
e^{-\l_0/(c_\lambda\l)}(c_\lambda\l/\l_0)^{2}\\
\label{aan}
\kappa(\l) & = \kappa_0 \biggl[1+ \kappa_1 \l +
\bar \kappa_0 \left(1+\frac{\bar \kappa_1 \l_0}{c_\lambda\l} \right) e^{-\l_0/(c_\lambda\l) }\frac{(c_\lambda\l/\l_0)^{4/3}}{\sqrt{\log(1+c_\lambda\lambda/\l_0)}}\biggr]^{-1}  \\
\label{wan} w(\l,\tau_i) & =  \frac{w_0}{1+ \beta_s \tanh({\gamma_s^2 \tau_i^2})}\biggl[1 + \frac{w_1 {c_\lambda}\l/\l_0}{1+ {c_\lambda}\l/\l_0}  + \bar w_0
e^{-\l_0/(c_w\l)}\frac{(c_w\l/\l_0)^{4/3}}{\log(1+c_w\lambda/\l_0)}\biggr]^{-1} \ .
\end{align}
The parameters of the glue function $\Vg$ are taken from~\cite{Jokela:2018ers}:
\begin{align}
    V_{g,1} = \frac{11}{27 \pi^2}\ , ~~~~ V_{g,2} = \frac{4619}{46656 \pi^4}\ , ~~~~ \lambda_0 = 8 \pi^2\ , ~~~~ V_\mathrm{IR} = 2.05\ , ~~~~ c_{\lambda}=3\ .
\end{align}
The parameters of the remaining functions $\Vf$, $\kappa$, and $w$ are taken from the finite strange quark-mass variant of V-QCD model (the ``standard potentials'' in this reference)~\cite{Jarvinen:2025mgj}. The Tables~\ref{tab:Pota} and~\ref{tab:Potb} below show the parameters for the potential fitted to lattice data. Here the radius of curvature is given by
\be
 \ell = \sqrt{\frac{1}{1 - W_0/12}}\ .
\ee

\begin{table}[H]
\centering
\begin{tabular}{|c|c|c|c|c|c|c|c|c|}
\hline
Parameter &$W_0$ & 
$\Lambda/$MeV & $W_\mt{IR}$ & $\bar{\kappa}_0$ & $ \bar{\kappa}_1$
& $45 \pi^2 M^3 \ell^3 /\left( 1+7/4 \right)$  \\
\hline
Value & 5.886 & 
158.155 & 1 & 3.35 & 0.2
& 1.22   \\
\hline
\end{tabular}
\caption{Fit to thermodynamics for $\mu = 0$. Here $\ell$ is the UV AdS radius and $M$ is the five-dimensional Planck mass normalized to give the Stefan--Boltzmann law for the pressure at higher temperature. $\Lambda$ is the energy scale of the background solution (see Appendix~\ref{app:asymptotics}).}
\label{tab:Pota}
\end{table}

\begin{table}[H]
\centering
\begin{tabular}{|c|c|c|c|c|c|c|c|c|c|}
\hline
Parameter & $w_0$ & $c_w$ & $\bar{w}_0$  & 
$w_1$ & 
$\beta_s$ & $ \gamma_s$ \\
\hline
Value & 0.81 & 1.1 & 12 & 
0.4  
&0.65 & 10   \\
\hline
\end{tabular}
\caption{Potential parameters obtained by fit to the light and strange quark susceptibility for $\mu = 0$.}
\label{tab:Potb}
\end{table}

In addition, the parameters $\kappa_0$ and $\kappa_1$ are given by
\begin{align}
    \kappa_0 = \frac{3}{2} - \frac{W_0}{8} \ , ~~~~ \kappa_1 = \frac{11}{24\pi^2}
\end{align}
and the parameters $W_1$ and $W_2$ are given by
\begin{align}
    W_1 = \frac{8+3 W_0}{ 9 \pi^2}\ , ~~~~ W_2 = \frac{6488+999 W_0}{15552 \pi^4}\ .
\end{align}
These parameters are not fitted to lattice data but their values are determined by comparing to the RG flow in QCD perturbation theory~\cite{Jarvinen:2011qe,Alho:2012mh}.

\section{Near-boundary asymptotics}\label{app:asymptotics}

Here we describe the near boundary asymptotics of the geometry and the tachyon. The higher order expansion for the geometry is used as it improves the numerical convergence for determining the near-boundary energy scale $\Lambda$ of the theory. This number is one of the sources in the geometry~\cite{Gursoy:2007cb,Gursoy:2007er}, and therefore needs to be kept constant at high precision when computing the numerical backgrounds.

The near-boundary asymptotics of the geometry depend on the weak-coupling expansion of the effective potential~\cite{Jarvinen:2011qe} 
\begin{equation}
\Vg(\lambda) - x \Vf(\lambda,0) = \frac{12}{\ell^2}(1+v_1\lambda+ v_2\lambda^2+v_3\lambda^3+v_4\lambda^4+ \mathcal{O}(\lambda^5))\ , 
\end{equation}
where $\ell$ is the radius of curvature. In terms of $v_i$, the scale factor and dilaton have the following near boundary expansions:
\begin{align}
\label{eq:Aexp}
A(r) =& -\log\frac{r}{\ell}
+ \frac{4}{9\log(r\Lambda)} \nonumber \\
&+ \frac{
\left(\frac{95}{162}-\frac{32v_2}{81v_1^2}\right)
+ \left(-\frac{23}{81}+\frac{64v_2}{81v_1^2}\right)
\log\left(-\log(r\Lambda)\right)
}{
\left(\log(r\Lambda)\right)^2
}
+ \mathcal{O}\left(\frac{1}{\left(\log(r\Lambda)\right)^3}\right) \\
v_1\lambda(r) =& -\frac{8}{9\log(r\Lambda)}
+ \frac{
\left(\frac{46}{81}-\frac{128v_2}{81v_1^2}\right)
\log\left(-\log(r\Lambda)\right)
}{
\left(\log(r\Lambda)\right)^2
}
+ \mathcal{O}\left(\frac{1}{\left(\log(r\Lambda)\right)^3}\right)\ .
    \label{eq:lambdaexp}
\end{align}
Combining these two leads to the small dilaton expansion for $A$
\begin{align}
    A(\lambda) =& \log(\Lambda \ell) +\frac{8}{9v_1 \lambda}-\left(\frac{23}{36}-\frac{16v_2}{9v_1^2} \right)\log\left(\frac{9}{8}v_1\lambda\right) \nonumber\\
    & +
    \left(-\frac{7v_1}{64}+\frac{25v_2}{18v_1}-\frac{32v_2^2}{9v_1^3}+\frac{8v_3}{3v_1^2} \right)\lambda \nonumber\\
    &\quad +\left(-\frac{119v_1^2}{8192}+\frac{31v_2}{128}+\frac{32v_2^3}{9v_1^4}+\frac{59v_3}{36v_1}-\frac{16v_2v_3}{3v_1^3}+\frac{32v_4-25v_2^2}{18v_1^2}\right)\lambda^2 + \mathcal{O}(\lambda^3) \ ,
\end{align}
where the first term is essentially an integration constant, and we included higher-order terms that cannot be directly obtained by using~\eqref{eq:Aexp} and~\eqref{eq:lambdaexp}. We use this second-order expansion in $\lambda$ to set the energy scale $\Lambda$ in our numerics.

The asymptotics of the tachyon depends on the expansions of $\kappa$ as well as the mass of the tachyon, obtained from expanding the potential $\Vf(\lambda,\tau_i)$ to second order in the tachyon. However, we choose the potentials such that this expansion is simply
\begin{equation}
\Vf(\lambda,\tau_i) = V_{\mt{f}\lambda}(\lambda)(1-\tau_i^2+\mathcal{O}(\tau_i^4))\ .
\end{equation}
In this case, we can define the expansion of $\kappa$ as 
\begin{equation}
\kappa(\lambda)=\kappa_0(1-\kappa_1\lambda +\mathcal{O}(\lambda^2)) \ .
\end{equation}
Consequently, the near-boundary expansion for the tachyon reads 
\begin{equation} \label{eq:tachyonBdry}
\tau_i(r) = m_i r \ell (-\log(r\Lambda))^\gamma \left(1+\mathcal{O}\left(\frac{1}{\log(r\Lambda)}\right)\right) \ ,
\end{equation}
where $\gamma = \frac{4}{3}-\frac{4\kappa_1}{3v_1}$. The factor of $\ell$ follows the normalization conventions in \cite{Jarvinen:2011qe,Alho:2012mh}.

\section{Computational details}\label{app:details}

Here we describe some computational details about evaluating the pressure (\ref{eq:pressureint}) and cumulants (\ref{eq:chindefunconst}).

\subsection{Computation of the background}
The black hole backgrounds are constructed numerically by solving the coupled Einstein-dilaton-tachyon equations following the flavor-dependent V-QCD action \eqref{eq:actionfull}. The complete background equations read,
\begin{align}
         {f}'' + 3 {A}' {f}' - \frac{ e^{2 A}}{\Nc} \sum_{i=1}^{\Nf} \frac{\sqrt{1+ e^{-2 A} f {\tau'}_i^2 \kappa}}{\sqrt{1 + K_i}} K_i V_{\mt{f}i} = 0
          \end{align}
          \begin{align}
    & 3 {A}'' + 3 {A'}^2 + 3 \frac{{A}' {f}'}{2 f} + \frac{2 {\lambda'}^2}{3 {\lambda}^2}- \frac{e^{2 A} \Vg}{2 f} + \frac{ e^{2 A}}{2 \Nc f} \sum_{i=1}^{\Nf} V_{\mt{f}i} \sqrt{1+ e^{-2 A} f {\tau'}_i^2 \kappa} \sqrt{1 + K_i} = 0 
    \end{align}
    \begin{align}
    & \frac{{\lambda''}}{\lambda} - \frac{{\lambda'}^2}{\lambda^2} + \frac{{f'} {\lambda'}}{f \lambda} + \frac{3 {A'} {\lambda'}}{\lambda} +\frac{3 e^{2 A}}{8 f} \lambda \partial_{\lambda} \Vg  & \\ 
    &- \frac{3}{8\Nc}\sum_{i=1}^{\Nf} \frac{ \left(V_{\mt{f}i}\sqrt{1+ e^{-2 A} f {\tau'}_i^2 \kappa} \right) \lambda  e^{2A}  }{ f \sqrt{1+ K_i}} \left(\partial_\lambda \log \left( V_{\mt{f}i} \sqrt{1+ e^{-2 A} f {\tau'}_i^2 \kappa}  \right) -  K_i \partial_\lambda \log w_i \right)  \nonumber   = 0 
    \end{align}
    \begin{align}
    & (1+K_i) {\tau''_i} + \left( {\tau'}_i^2 + \frac{e^{2 A}}{f \kappa}\right) \left( K_i \partial_{\tau_i} \log w_i - \partial_{\tau_i} \log V_{\mt{f}i} \right) + {\tau'}_i^3 {A'} e^{-2 A} f \kappa(1+K_i) \nonumber \\ & +  \left( {\tau'}_i {\lambda'} \left( \partial_\lambda \log V_{\mt{f}i} - K_i \partial_\lambda \log w_i \right) + 3 {\tau'}_i {A'} \right) \left( 1+ {\tau'}_i^2 e^{-2 A} f \kappa \right) \nonumber \\ & + (1+ K_i) {\tau'}_i \left({\kappa'} + {f'} \right)\left( 1+ \frac{1}{2}{\tau'}_i^2 e^{-2 A} f \kappa \right) = 0
\end{align}
\begin{align} \label{eq:constraint}
    \frac{{\lambda'}^{2}}{\lambda^2} + 9 {A'}^2 + \frac{9}{4} \frac{{A'} {f'}}{f} - \frac{3 e^{2 A}}{4 f}\left( \Vg - \frac{1}{\Nc} \sum_{i} V_{\mt{f}i} \frac{\sqrt{1+K_i}}{\sqrt{1+ e^{-2 A} f \tau_i'^2 \kappa}}\right) = 0 
    \end{align}
where
\begin{equation}
    V_{\mt{f}i} \equiv \Vf(\lambda,\tau_i) \ , \qquad w_i \equiv w(\lambda,\tau_i) \,  \qquad i = u,d,s\ .
\end{equation}
and primes denote differentiation with respect to the original holographic radial coordinate $r$. $K_i$ encode the contribution from the flavor charge densities $\bar{n}_i$ and is defined as,
\begin{align}
    K(\lambda, \tau_{i},\bar n_i,A) \equiv K_{i} =  \frac{\left( e^{-3 A}  \bar  n_i \right)^2}{\left( w(\lambda,\tau_{i})\Vf(\lambda,\tau_{i})\right)^{2}} \ .
\end{align}
For the $2+1$-flavor solutions, the light-quark masses are set to zero, $m_u=m_d=0$, which on the chirally symmetric black-hole branch, refers to the vanishing of the corresponding tachyon profiles, $\tau_u(r)=\tau_d(r)=0$.
The strange quark tachyon, in contrast, remains nonzero due to the finite strange-quark mass, whose horizon value $\tau_{s,h}\equiv\tau_s(r_h)$, is therefore treated as an additional parameter.

For the numerical integration, instead of radial coordinate $r$, it is convenient to use the scale factor $A$ itself as the radial variable~\cite{Alho:2012mh,Alho:2013hsa}. Here the black-hole horizon is set at $A_h\equiv A(r_h)=0$ using the scaling symmetry of the background equations, and the integration is initialized slightly outside the horizon at $A=\epsilon$, using the regular near-horizon expansions. The solutions are parametrized by the horizon coupling $\lambda_h\equiv\lambda(r_h)$,
together with the horizon values of the tachyon fields $\tau_{s,h}$, and the dimensionless flavor densities\footnote{Note that our notation differs from that used in~\cite{Jarvinen:2025mgj} by a factor of $\Nc$.} $\tilde n_i = e^{-3 A_\mt{h}}\bar n_i/\Nc$. Once these quantities are specified, regularity at the horizon fixes the remaining coefficients in the near-horizon expansion and thereby provides the initial data for the numerical integration.

With this initial data, the coupled background equations are integrated from the near-horizon region towards the ultraviolet region in two steps. At first the full background system is evolved up to an intermediate cut-off $A_{\mathrm{UV}}^{(1)}$ where the solution is already deep in the ultraviolet regime. The integration is then continued using the ultraviolet form of the equations, in particular for the tachyon fields, up to the second cut-off $A_{\mathrm{UV}}^{(2)}$.
This two-step procedure allows the solution to approach sufficiently closely the asymptotic regime for the ultraviolet parameters to be extracted reliably.

The energy scale $\Lambda$ is determined by matching the numerical solution for $A$ and $\lambda$ to the ultraviolet expansions derived in Appendix~\ref{app:asymptotics}. In practice, we use the expansion of $A(\lambda)$ through second order in $\lambda$ given above. Since the additive constant in $A(\lambda)$ contains $\log(\Lambda\ell)$, comparison of the numerical solution with this asymptotic form determines the energy scale $\Lambda$ associated with each background. The strange quark mass extracted from the leading non-normalizable tachyonic mode from \eqref{eq:tachyonBdry} is kept fixed by tuning the horizon value of the $\tau_{s,h}$. Following~\cite{Jarvinen:2025mgj}, we set the strange quark mass to $m_s/\Lambda = 0.2839$. This value is lower than in~\cite{CruzRojas:2024etx}, where the value was obtained by fitting kaon masses, since a non-zero quark mass introduces non-linearities in the tachyon, which enhances with increasing mass, making precise fitting of the data challenging at low temperature. 

The remaining horizon and density parameters are determined according to the physical condition imposed on the system as follows.
In the $\beta$-equilibrium setup, charge neutrality and chemical equilibrium condition requires $\nQ=0=2 n_u/3 -n_d/3-n_s/3$, and $\mu_d = \mu_s$. While implementing these conditions numerically, we parametrize the densities like $\tilde{n}_d = \tilde{n}_u+\delta \tilde{n}$ and $\tilde{n}_s = \tilde{n}_u-\delta \tilde{n}$ and pass it to the background solver. Thus, the $\delta\tilde{n}$ controls the relative strange-quark density, and it is varied until $\mu_d = \mu_s$. For a prescribed target temperature, $\lambda_h$ is first varied until the numerical solution gives $ T=T_{\rm goal}$. Then $\tau_{s,h}$ is adjusted until the ultraviolet strange-quark mass is satisfied. Finally, the relative flavor-density parameter is varied until the chemical-equilibrium condition $\mu_d=\mu_s$ is satisfied. These steps are repeated using root searches until the temperature, strange-quark mass, and chemical-potential difference simultaneously satisfy the required numerical tolerance. In this sense, the three shooting parameters play the schematic roles
\begin{align}
    \lambda_h \rightarrow T\ ,\qquad \tau_{s,h}\rightarrow \frac{m_s}{\Lambda}\ ,\qquad \delta\tilde n\rightarrow \mu_d-\mu_s \ .
\end{align}

For the background relevant for HIC different flavor composition is imposed. The requirement of strangeness neutrality imposes $ n_s=0$ and the phenomenological HIC condition requires $\nQ/\nB=\frac{2}{5}$. These conditions leads to following parameterization of the dimensionless densities, $\tilde{n}_s = 0$ and
$\tilde{n}_d = 8 \tilde{n}_u/7$, which is passed to the background solver. The same numerical background solver is then used, with $\tau_{s,h}$ adjusted to maintain the fixed strange quark mass and the remaining horizon parameters varied to obtain the desired thermodynamic state. Once the background has been determined, the flavor chemical potentials, susceptibilities and pressure are obtained by integrating the corresponding gauge field equations on that background.

\subsection{Computing the pressure and cumulants}\label{app:pandsusc}

To obtain the pressure, we numerically integrate 
\begin{equation}
    \dd P = s\dd T + \nB \dd \muB+\nQ\dd\muQ+n_\mathrm{S}\dd\muS \ .
\end{equation}
The flavor structure on the right hand side can be further reduced in the various constrained setups. To obtain the pressure at finite density, we split the integration to two parts. The zero-density contribution to the pressure is obtained by integrating $sdT$. This gives
\begin{equation}
P(T,\muB=0) = \int_{T^*}^T\dd\tilde{T}s(\tilde{T},\muB=0) \ ,
\end{equation}
where $T^*$ is the confinement temperature at which the pressure becomes zero. This temperature can be computed by following~\cite{Jarvinen:2025mgj}. To obtain the pressure at finite density, we then integrate 
\begin{equation}\label{eq:pressureint}
P(T,\muB) = P(\bar{T},\muB=0) + \int_\gamma \dd P \ ,
\end{equation}
where $\gamma(\rho), \ \rho \in [0,1]$ is a path in the $(\muB,T)$-plane such that $\gamma(0) = (0,\bar{T})$ and $\gamma(1) = (\muB,T)$. The dependence on the intermediate temperature $\bar{T}$ cancels from the final result because $\dd P$ is an exact differential. One could perform the finite-density integration at fixed $T$. However, in these models, the natural control variable for the temperature is the horizon value of the dilaton $\lambda_h$, but constant $\lambda_h$ contours do not correspond to fixed temperature in general.

For the $\beta$-equilibrium setup we require charge neutrality $n_\mathrm{Q}=0$ and $\mu_d=\mu_s$. The first law becomes
 \begin{align}
    \dd P &= s \dd T + n_u \dd \mu_u+n_d \dd \mu_d + n_s \dd\mu_s  \\
     &= s \dd T + n_u(\dd\mu_u + \dd\mu_d + \dd\mu_s) = s\dd T +\nB\dd\muB \ . 
 \end{align}
 For the HIC setup we require zero strangeness $n_\mathrm{S}=0$ and fix the ratio of charge to baryon density to $\nQ/\nB = 0.4$. The conditions give us only one free quark density, which we take to be $n_u$. 
 For general fixed ratio $\nQ/\nB =\alpha$ we get $ n_d = (2-\alpha)n_u/(1+\alpha)$, which further gives $\nQ = \alpha n_u/(1+\alpha)$ and $\nB = n_u/(1+\alpha)$.
The first law for this case reads 
\begin{align}
\dd P &= s \dd T + n_u \dd\mu_u+n_d \dd\mu_d + n_s \dd \mu_s \\ &= s \dd T + n_u\left(\dd\mu_u+\frac{2-\alpha}{1+\alpha}\dd\mu_d\right) = sdT + \nB(\dd\muB+\alpha \dd\muQ) \ .
\end{align}

In both models we parametrize the finite density phase diagram points via the dimensionless density parameter $\tilde{n}_i$ which is defined as
 \begin{equation}
   \tilde{n}_i = n_i \frac{4\pi}{s}\ .   
 \end{equation} 
After imposing the constraints we are left with only one independent density parameter in both $\beta$-equilibrium and HIC setups. We choose it to be the dimensionless density parameter for the up-quark $ \tilde{n}_u$. The other independent parameter is the horizon value for the dilaton $\lambda_h$, which roughly controls the temperature. The points in the phase diagram we can thus parametrize in terms of $ \tilde{n}_u$ and $\lambda_h$. To evaluate the finite density pressure contribution in (\ref{eq:pressureint}) we choose the path $\gamma=(\muB(\tilde{n}_u,\lambda^\star_h),T(\tilde{n}_u,\lambda^\star_h))$ with $\lambda^\star_h$ fixed to give the temperature $\bar{T}$ in (\ref{eq:pressureint}) at $\tilde{n}_u = 0$.
 For the $\beta$-equilibrium setup we get the pressure contribution
  \begin{equation}
    \int_\gamma \dd P = \int _0^{\tilde{n}^*_u} \left[s\frac{\dd T}{\dd\tilde{n}_u} +n_u \frac{\dd\muB}{\dd\tilde{n}_u}\right]\dd\tilde{n}_u \ ,
 \end{equation}
 and for the HIC setup we get
 \begin{equation}
    \int_\gamma \dd P = \int _0^{\tilde{n}^*_u}\left[ s\frac{\dd T}{\dd\tilde{n}_u} + n_u\left(\frac{\dd\mu_u}{\dd\tilde{n}_u}+\frac{2-\alpha}{1+\alpha} \frac{\dd\mu_d}{\dd\tilde{n}_u}\right) \right]\dd\tilde{n}_u \ ,
 \end{equation}
 where $\tilde{n}^*_u$ is the end point of the integration. The pairs $(\lambda^\star_h,\tilde{n}^*_u)$ therefore provide a parametrization for the values  $(\muB,T)$ on the phase diagram.

Next we describe how we evaluate the ratios $C_n/C_m$ of the cumulants
(\ref{eq:hiccum}). In the constrained setups the electric charge and strangeness chemical potentials are functions
of $\muB$, determined by the conditions imposed on the densities, so
that the pressure reduces to an effective function
$P_\text{eff}(T,\muB)=P(T,\muB,\muQ(\muB),\muS(\muB))$ of two
variables. Differentiating $P_\text{eff}$ with respect to $\muB$ then
produces the chain-rule terms
\begin{equation}\label{eq:neff}
    \frac{1}{T^3}\frac{\partial P_\text{eff}(T,\muB)}{\partial \muB}
    = \frac{\nB}{T^3}
    + \frac{\nQ}{T^3} \frac{\dd\muQ}{\dd\muB}
    + \frac{\nS}{T^3}\frac{\dd\muS}{\dd\muB}
    = \frac{n_\text{eff}}{T^3} \ ,
\end{equation}
so that the constrained susceptibilities generated by $P_\text{eff}$
differ from the standard ones. The cumulants compared to heavy-ion collision data are instead evaluated at fixed
$\muQ$ and $\muS$, which is the definition (\ref{eq:chindefunconst})
used on the lattice. Concretely, for each freeze-out point $k$ we first
read off the values $\muQ^0$ and $\muS^0$ attained on the HIC-constrained
trajectory at $(T^0,\muB^0)$, and we then hold these two chemical
potentials fixed while varying $\muB$ and $T$. Because $\muQ$ and
$\muS$ are constants throughout a batch, $\dd\muQ=\dd\muS=0$ and the
first law collapses to
\begin{equation}
\dd P = s\,\dd T + \nB\,\dd\muB \ ,
\end{equation}
so that the two chain-rule terms in (\ref{eq:neff}) are absent and
$n_\text{eff}=\nB$ identically. The derivatives
we compare to heavy-ion data are the standard baryon-number susceptibilities
(\ref{eq:chindefunconst}) rather than the constrained ones.
 
To evaluate the cumulants at the freeze-out points we generate two-dimensional batches of data centred at each of the
freeze-out points of Fig.~\ref{fig:diagram}, each batch computed at its
own fixed $(\muQ^0,\muS^0)$. We do this by defining a grid of $\lambda_h$
and $\tilde n_u$ centered at $\lambda^0_h$
and $\tilde n^0_u$ which are defined by giving the centers of the freeze-out points along the HIC contrained trajectory. At each grid point we then solve for the remaining
two densities $\tilde n_d$ and $\tilde n_s$ from the two conditions
\begin{align}\label{eq:qsconditions}
&\mu_u(\tilde n_u,\tilde n_d,\tilde n_s;\lambda_h)-\mu_d(\tilde n_u,\tilde n_d,\tilde n_s;\lambda_h) = \muQ^0 \ , \nonumber \\
&\mu_d(\tilde n_u,\tilde n_d,\tilde n_s;\lambda_h)-\mu_s(\tilde n_u,\tilde n_d,\tilde n_s;\lambda_h) = \muS^0 \ .
\end{align}
Since the model has vanishing off-diagonal quark susceptibilities, each
chemical potential is to a very good approximation a function of its own
density alone, $\mu_i \simeq \tilde n_i\, I_i[\lambda_h]$ with $I_i$
varying only weakly with the densities. The strange quark mass
is held at its physical value throughout by solving for the horizon
tachyon $\tau_{h}$ at every point. The baryon density is then
\begin{equation}
    \nB = \frac{s}{4\pi}\,\frac{\tilde n_u+\tilde n_d+\tilde n_s}{3} \ .
\end{equation}
 
To these batches we fit a polynomial
\begin{equation}\label{eq:polyfit}
\text{Poly}_m[\nB/T^3](\muB,T)
= \sum_{i=0}^m\sum_{j=0}^{m-i} a_{ij}
\left(\frac{\muB-\muB^0}{T}\right)^i
\left(\frac{T-T^0}{\Delta T}\right)^j \ ,
\end{equation}
where $(\muB^0,T^0)$ is the freeze-out point and $\Delta T$ is the
half-width of the batch in the temperature direction. Both expansion variables are dimensionless, so the coefficients
$a_{ij}$ are dimensionless as well. From this fit we define the ratios
\begin{equation}
\mathrm{Func}_m[C_k/C_s](\muB,T)
=
\frac{
\left(\dfrac{\partial}{\partial(\muB/T)}\right)^{k-1}
\,\mathrm{Poly}_m[\nB/T^3](\muB,T) \bigg{\lvert}_{T,\,\muQ,\,\muS \ \text{fixed}}
}{
\left(\dfrac{\partial}{\partial(\muB/T)}\right)^{s-1}
\,\mathrm{Poly}_m[\nB/T^3](\muB,T)  \bigg{\lvert}_{T,\,\muQ,\,\muS \ \text{fixed}}
} \ ,
\end{equation}
and evaluate them at the centre of the batch,
\begin{equation}
C_k/C_s = \text{Func}_m[C_k/C_s](\muB^0,T^0) \ .
\end{equation}

The dominant uncertainty in Fig.~\ref{fig:qgpcumulants} comes from the freeze-out points themselves, which we propagate as
\begin{equation}
    \sigma_\text{fo} = \sqrt{
    \left(\partial_{\muB}\text{Func}_m[C_k/C_s](\muB^0,T^0)\,\delta \muB\right)^2
    +\left(\partial_T\text{Func}_m[C_k/C_s](\muB^0,T^0)\,\delta T\right)^2 } \ ,
\end{equation}
where $\delta \muB$ and $\delta T$ are the standard deviations of
$\muB$ and $T$ shown in Fig.~\ref{fig:diagram}, which come from the
uncertainty of the mapping $\sqrt{s_\mathrm{NN}} \leftrightarrow
(\muB,T)$ \cite{STAR:2017sal}. The derivatives are taken with the other
variable held fixed. In addition we quantify the sensitivity of the result to the fitting procedure itself. For each batch we repeat the extraction over an
ensemble of fits, varying the degrees and the fraction of the batch included in the
fit window. The quoted value is the median and the fit systematic error $\sigma_\text{fit}$ is its half-spread. Across the ensemble the ratios are stable at the level of a few tenths of a percent for $C_2/C_1$ and $C_3/C_2$ and
below one percent for $C_4/C_2$ at the higher collision energies,
growing to a few percent at the lowest energies. Since
$\sigma_\text{fit}$ is everywhere small compared with $\sigma_\text{fo}$,
the total error $\sigma=\sqrt{\sigma_\text{fo}^2+\sigma_\text{fit}^2}$
shown in Fig.~\ref{fig:qgpcumulants} is dominated by the freeze-out
uncertainty.

We also display some data for the standard baryon number susceptibilities at zero density
\begin{equation}
\chitwo = \frac{1}{T^2}\frac{\partial^2 P(T,\muB,\muQ,\muS)}{\partial\muB^2}\bigg{\lvert}_{\muB=0 , \ \muQ=0 , \ \muS=0}
\end{equation}
in Fig.~\ref{fig:chiaux} (right), which are compared to lattice results. To calculate these we parametrize $n_d = n_u$ and $n_s = n_u-\delta n$ and tune $\delta n$ so that $\mu_d=\mu_s$. Because our model produces zero off-diagonal quark susceptibilities ($\chi^{ij}=0$ with $i\neq j$) and up-and down-quarks are degenerate in mass, the choice $n_d = n_u$ gives $\mu_d=\mu_u$. In the conserved charge basis $\mu_u=\mu_d$ and $\mu_d=\mu_s$ then imply $\muQ = \muS=0$. With this setup we then generate data with small $n_u$ which gives small $\muB$. Then the baryon number susceptibilities are obtained by numerically differentiating the baryon number density
\begin{equation}
    \nB = \frac{s}{4\pi}(\tilde{n}_u-\frac{\delta \tilde{n}}{3})
\end{equation}
with respect to $\muB$. We also display the quark susceptibilities in Fig.~\ref{fig:chiaux} (left) that are used in the fit, for which we follow \cite{Jarvinen:2025mgj}. The results of the HotQCD group \cite{HotQCD:2012fhj} do not display the light quark susceptibility $\chi^{ll}$ explicitly. By using (\ref{eq:chi2Btransf}) we find $\chi^{ll}=(9\chi_2^{\mathrm{B}}-\chi^{ss}-2\chi^{ud}-4\chi^{ls})/2$ which combined with the results of \cite{HotQCD:2012fhj} yield the light quark susceptibility for the HotQCD group.

\section{Comparison with lattice data and the offset of higher-order baryon number cumulants} \label{app:cumulants}

In this appendix, we detail the phenomenological reweighting procedure used to address a specific limitation of our holographic setup. Because the V-QCD action relies on a diagonal DBI flavor action, it lacks direct multi-trace couplings between distinct flavor sectors, yielding vanishing off-diagonal quark susceptibilities at zero density. Here we explain how this omission leads to the discrepancy between the heavy-ion data and our V-QCD results for the cumulants in the HIC setup shown in Fig.~\ref{fig:qgpcumulants}. 

The deviations in Fig.~\ref{fig:qgpcumulants} come from the fact that our V-QCD model (\ref{eq:Sfflavors}) produces zero off-diagonal quark susceptibilities ($\chi^{ij}=0$ with $i\neq j$). The cumulants are susceptible to the cross correlation between quarks as can be seen from the  baryon susceptibilities
\begin{align}
\chitwo &= \frac{1}{T^2}\frac{\partial^2 P(T,\muB,\muQ,\muS)}{\partial\muB^2} \\ &= \frac{1}{9}\left( \chi^{uu}+\chi^{dd}+\chi^{ss} + 2\chi^{ud}+2\chi^{us}+2\chi^{ds}\right) \ , 
\end{align}
where $\chi^{ij} = T^{-2}\partial^{2} P/\partial\mu_i\partial\mu_j$ with $i,j\in \{u,d,s\}$ are the quark susceptibilities. The formula follows readily by using 
\begin{align}
\left.\frac{\partial}{\partial \muB}f \right |_{\muQ,\muS \text{ fixed}}
=\frac{1}{3}
\left(
\frac{\partial }{\partial \mu_u}
+
\frac{\partial }{\partial \mu_d}
+
\frac{\partial }{\partial \mu_s}
\right) f
\ ,
\end{align}
where $f$ is a test function and $T$ is also kept fixed while taking the derivatives. 

Our model cannot produce the off-diagonal susceptibilities $\chi^{ij}$ with $ i\neq j$.
To incorporate some effects of the cross-correlation at finite density we can reweight the cumulants by multiplying them by $\chi^\text{lat}_n(T)/\chi^\text{mod}_n(T)$, where $\chi^\text{lat}_n(T)$ is based on the lattice results\footnote{The function $\chi^\text{lat}_n(T)$ can be defined to be an interpolation of central values obtained by combining results of various collaborations, or a fit to the lattice data using a simple Ansatz. We use here an interpolation of the lattice results of \cite{HotQCD:2012fhj,Bazavov:2020bjn} for $\chi_2^\mathrm{B}$ and $\chi_4^\mathrm{B}$. As we can see from Fig.~\ref{fig:diagram} the relevant temperature range where we need these cumulants is roughly $T\in[140,170]$ MeV.} for $\chi_n^\mathrm{B}$ at $\muB=0$
and $\chi^\text{mod}_n(T)$ are the predictions of our model for the baryon susceptibilities at $\muB=0$. Both of these are evaluated at $\muB=0$. Also, it might be necessary to emphasize that these are the standard cumulants, where we are varying $\muB$ while keeping $\muQ=0$ and $\muS=0$ fixed. The evaluation of these is discussed in the end of Appendix~\ref{app:pandsusc}. As the odd cumulants are zero at $\muB=0$, for reweighting we instead use  $\chi^\mathrm{B}_1 \approx \chi^\mathrm{B}_2 \frac{\muB}{T}$ and $\chi^\mathrm{B}_3 \approx \chi^\mathrm{B}_4 \frac{\muB}{T}$. Because we are only expanding these to lowest order the chemical potential dependence cancels in the reweighting factor. If we use this method for  $C_2/C_1$, $C_3/C_2$, and $C_4/C_2$ we obtain the reweighted cumulant ratios \begin{equation}\label{eq:reweightrat}
\frac{C_2}{C_1} \rightarrow\frac{C_2}{C_1}\ , \quad \frac{C_3}{C_2} \rightarrow \frac{\chi^\text{lat}_4(T)/\chi^\text{mod}_4(T)}{\chi^\text{lat}_2(T)/\chi^\text{mod}_2(T)}\frac{C_3}{C_2}\ , \quad \frac{C_4}{C_2}  \rightarrow \frac{\chi^\text{lat}_4(T)/\chi^\text{mod}_4(T)}{\chi^\text{lat}_2(T)/\chi^\text{mod}_2(T)}\frac{C_4}{C_2}\ .
\end{equation}

\begin{figure}[!htb] 
    \centering
    \includegraphics[width=0.45\textwidth]{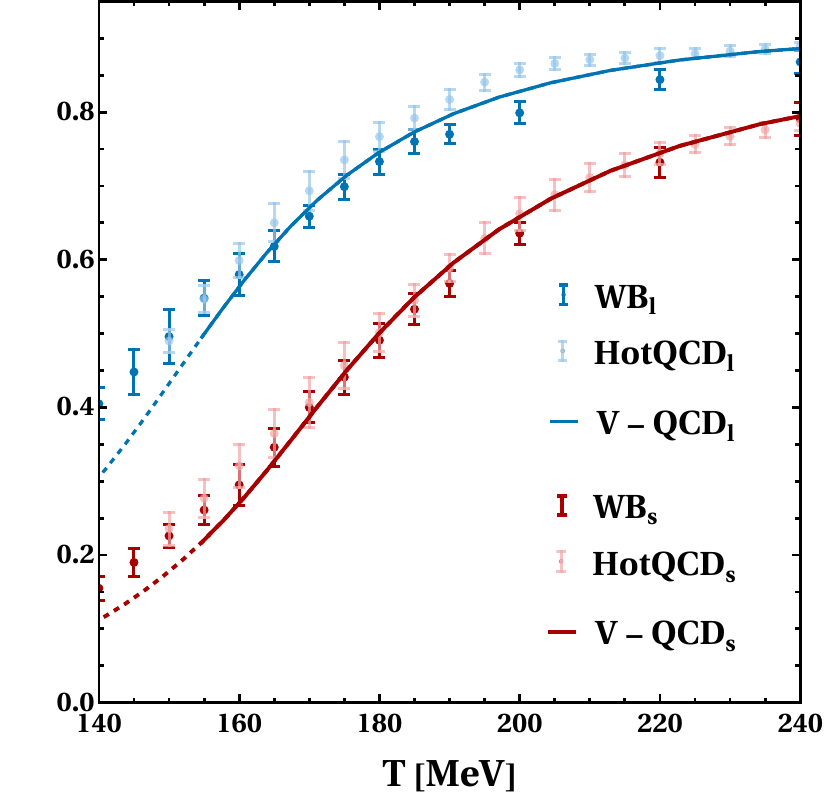}
    \includegraphics[width=0.425\textwidth]{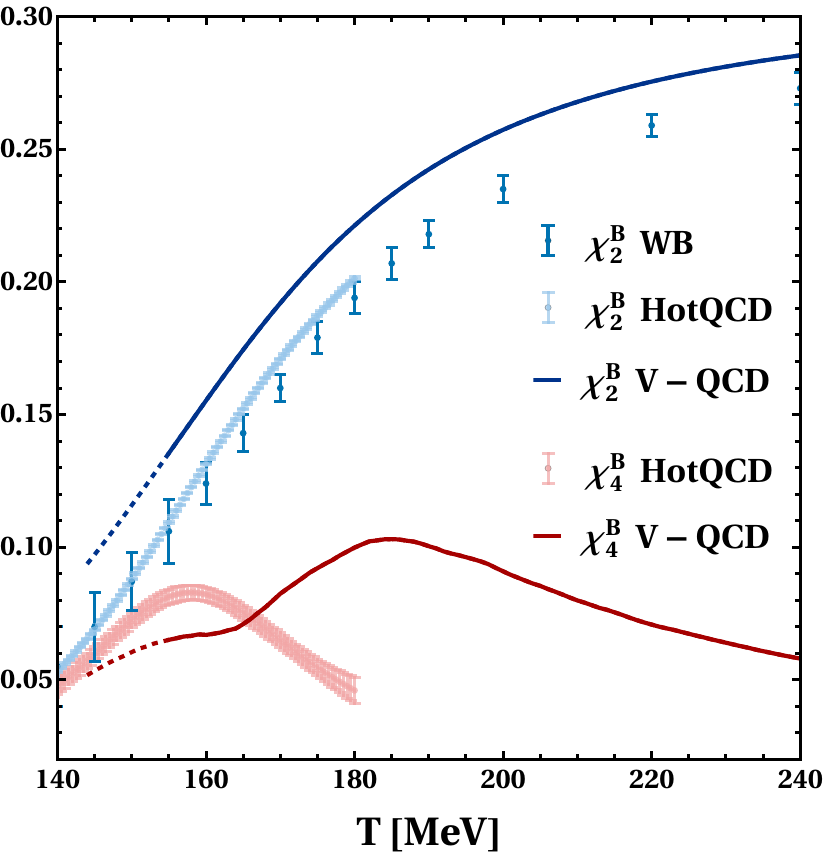}
    \caption{{\bf Left:} The light-quark ($l$) and strange-quark ($s$) susceptibilities as functions of temperature in the V-QCD model, compared with the corresponding WB lattice data \cite{Borsanyi:2011sw} and HotQCD lattice data \cite{HotQCD:2012fhj}. The fit of the V-QCD model to the light-quark and strange-quark susceptibilities was performed in \cite{Jarvinen:2025mgj}. The solid curve indicates the temperature range, $T>155$ MeV, used in the fit. Restricting the fit to this temperature range allows the susceptibilities to remain small at low temperatures, which in turn improves the matching of the high-density equation of state to nuclear matter equations of state. {\bf Right:} Comparison of our  $\chi^\mathrm{B}_2$ and $ \chi^\mathrm{B}_4$ as a function of temperature to lattice data. The WB data is from \cite{Borsanyi:2011sw} and the HotQCD data is from \cite{Bazavov:2020bjn}. The solid line shows the region that was fitted for the quark susceptibilities.  We see that around the temperatures of the freeze-out parameters of Fig.~\ref{fig:diagram} our $\chi^\mathrm{B}_2$ overestimates the lattice while $\chi^\mathrm{B}_4$ underestimates it. 
    }
    \label{fig:chiaux}
\end{figure}

The reweighting ratios in (\ref{eq:reweightrat}) are evaluated at the same temperatures as the freeze-out temperatures used to evaluate $C_n/C_m$. These freeze-out temperatures one obtains from Fig.~\ref{fig:diagram} by projecting the central values to the $\muB=0$ axis. In (\ref{eq:reweightrat}) we see that the reweighting scheme does not change $C_2/C_1$, which explains the good agreement in Fig.~\ref{fig:qgpcumulants}, but $C_3/C_2$ and $C_4/C_2$ are modified both by the same factor. In Fig.~\ref{fig:chiaux} we show the lattice data of \cite{Borsanyi:2011sw,Bazavov:2020bjn} and the results for our model for $\chi^\mathrm{B}_2$ and $\chi^\mathrm{B}_4$. As seen in the right panel of Fig.~\ref{fig:chiaux}, around the relevant freeze-out temperatures, our model overestimates $\chitwo$ relative to the lattice data while underestimating $\chifour$. Consequently, the ratio of these correction factors is greater than one, yielding $\frac{\chi^\text{lat}_4(T)/\chi^\text{mod}_4(T)}{\chi^\text{lat}_2(T)/\chi^\text{mod}_2(T)}>1$.

\begin{figure}[!htb]
    \centering
    \includegraphics[width=0.49\textwidth]{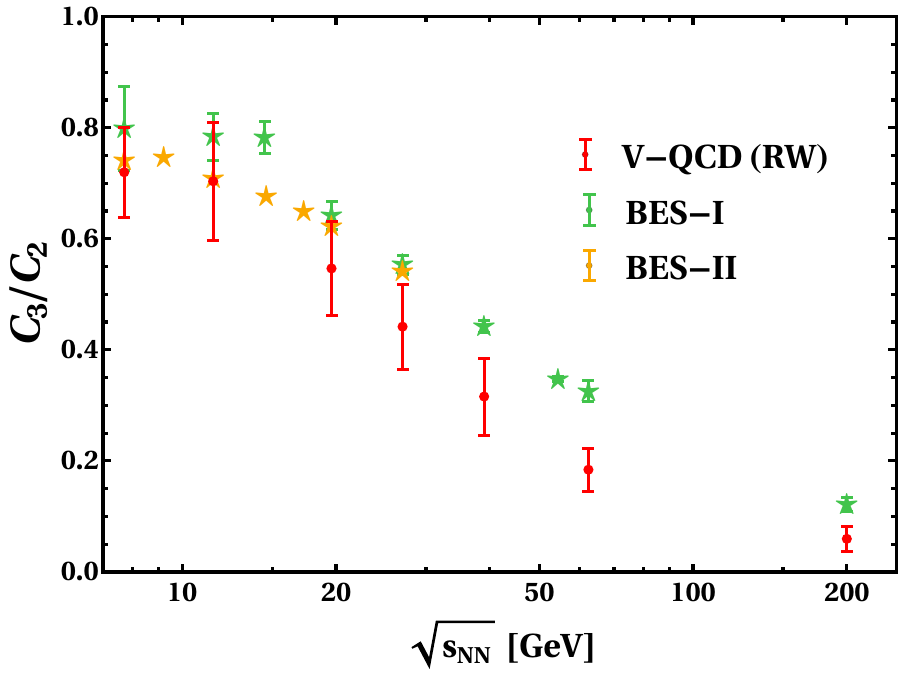}
    \includegraphics[width=0.5\textwidth]{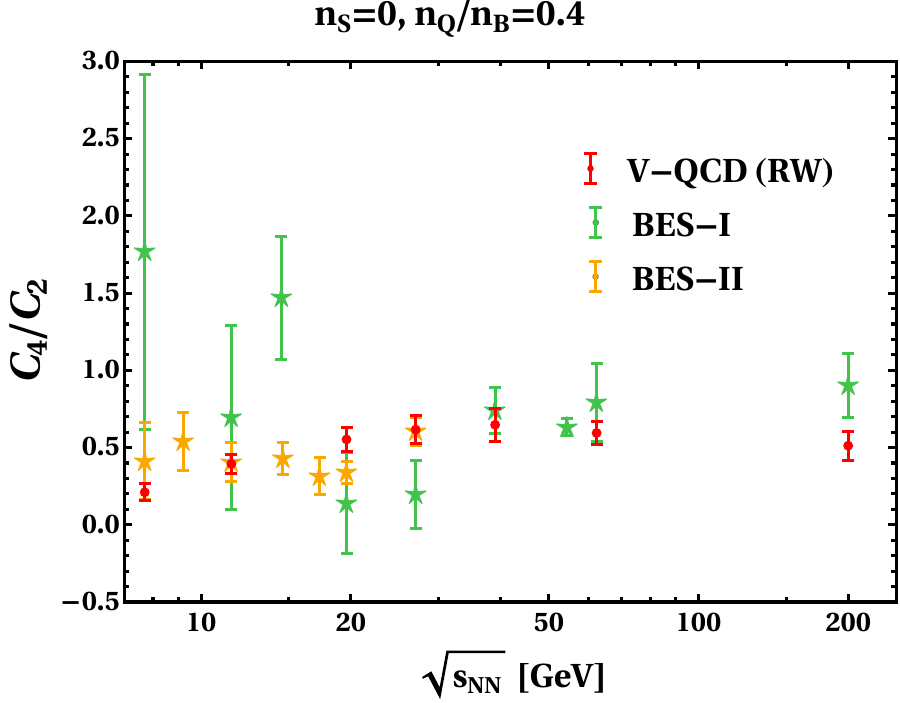}
    \caption{The reweighted (RW) cumulant ratios (\ref{eq:reweightrat}) are shown as a function of the collision energy $\sqrt{s_\mathrm{NN}}$. We show only $C_3/C_2$ and $C_4/C_2$, since the reweighting affects only these ratios. The results should be compared with those shown in Fig.~\ref{fig:qgpcumulants}. We find that the reweighting brings the predictions into closer agreement with the BES data.}
    \label{fig:qgpcumulantsrw}
\end{figure}

By examining the numerical values of the ratios $\chi^\text{lat}_2(T)/\chi^\text{mod}_2(T)$ and $\chi^\text{lat}_4(T)/\chi^\text{mod}_4(T)$, we find that our reweighting scheme for the cumulant ratios $C_3/C_2$ and $C_4/C_2$ can partially account for the discrepancy between our predictions and the heavy-ion data shown in Fig.~\ref{fig:qgpcumulants}. The results obtained after reweighting are shown in Fig.~\ref{fig:qgpcumulantsrw}. In general, the reweighting brings the predictions into closer agreement with the BES data, although complete agreement is not achieved. Several effects may contribute to the remaining discrepancy. These include uncertainties in the lattice data and freeze-out parameters and the dependence of the off-diagonal quark susceptibilities on the baryon chemical potential. Since these reweighting factors were introduced to phenomenologically incorporate the effects of nonzero non-diagonal quark susceptibilities, their success in improving the agreement with the data provides further evidence that the discrepancy is likely a consequence of neglecting these susceptibilities in the model.

\section{{Thermal-FIST for HRG model}}\label{app:thermalfist}

In Sec.~\ref{sec:heavyion}, we compare the V-QCD results with those obtained from two interacting hadron resonance gas models: the excluded-volume HRG model, EVX \cite{Rischke:1991ke, Vovchenko:2020lju}, and the quantum van der Waals HRG model, vdW \cite{Vovchenko:2015vxa, Vovchenko:2016rkn}. The thermodynamic quantities for both HRG models are computed using the \texttt{Thermal-FIST} package \cite{Vovchenko:2019pjl}. In particular,  we use the \texttt{QtThermalFIST} module, which provides a graphical interface (GUI) to the Thermal-FIST library and allows us to construct the corresponding equation of state. We specify the attractive interaction parameter $a$ and the short-range repulsive/excluded volume parameter $b$ following \cite{Vovchenko:2016rkn}. For the EVX model, we set $a=0$ and $b=3.42 ~\rm{fm}^3$ while for vdW we set $a = 329~\rm{MeV}\rm{fm}^3$ and $b=3.42~ \rm{fm}^3 $. The computations are done in the grand-canonical ensemble with constant Breit--Wigner width and the hadrons and resonances included in the PDG2014 particle list.

\begin{figure}[H]
\centering
\includegraphics[width=0.5\linewidth]{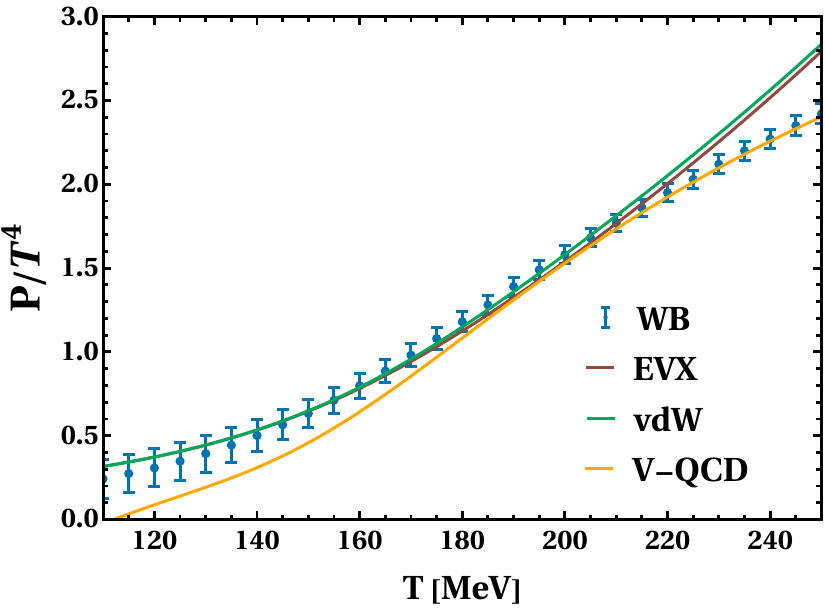}
\caption{Comparison of our V-QCD model to the EVX and vdW HRG models at $\muB =0$. The WB lattice data is from \cite{Borsanyi:2013bia}.}
    \label{fig:2}
\end{figure}

As a first step, we compare the HRG models with the V-QCD results at zero baryon chemical potential $\muB = 0$. Fig.~\ref{fig:2} compares the $P/T^4$ computed from the V-QCD model at $\muB=0$ with the EVX and vdW HRG models. We see that the V-QCD model intersects the EVX model around 180--200 $\text{MeV}$ but does not intersect the vdW model at $\muB=0$. This is consistent with the phase transitions observed, for example, in Fig.~\ref{fig:pdHRGbeta}. 

We then extend the comparison to finite baryon chemical potential in order to construct a more complete description of the dense QCD phase diagram by connecting the high-temperature V-QCD equation of state to a low-temperature hadronic description. In the low-temperature hadronic regime, interacting HRG models provide a benchmark for comparing the two physical realizations of the V-QCD equation of state: charge-neutral $\beta$-equilibrated matter and strangeness-neutral matter subject to the conditions relevant for HIC. These two sets of physical environments lead to four different HRG setups, corresponding to the two sets of constraints applied separately to EVX and vdW HRG models. In the charge-neutral $\beta$-equilibrium, both the HRG models are set such that the charge density is constrained to vanish, $\nQ=0$ while the strangeness chemical potential is set to zero, $\muS =0$, rather than being determined from the condition of zero net strangeness. In the second case, the strangeness-neutrality is imposed through $\nS = 0$ along with fixed charge-to-baryon density ratio, $\nQ/\nB = 0.4$. Within each of these setups, the HRG models are used as the low-temperature reference against the high temperature V-QCD branch to identify the phase boundary between the hadronic and deconfined phase via pressure matching under the corresponding thermodynamic constraints.

\bibliographystyle{JHEP}
\bibliography{refs} 

\end{document}